\documentclass[%
superscriptaddress,
preprint,
nofootinbib,
prd
]{revtex4-2}

\usepackage{filecontents}
\usepackage{subcaption}
\usepackage{float}
\usepackage{simpler-wick}
\usepackage{graphicx}
\usepackage{dcolumn}
\usepackage{bm}
\usepackage{amsmath}
\usepackage{dsfont}
\usepackage{siunitx}
\usepackage[english]{babel}
\usepackage{amsfonts}
\usepackage{braket,slashed,bm}
\usepackage{cancel}
\usepackage{standalone} 
\usepackage{tikz} 
\usepackage[normalem]{ulem}
\usepackage{hyperref}
\usepackage{multirow}
\hypersetup{
    colorlinks=true,
    linkcolor=blue,
    filecolor=blue,      
    urlcolor=blue,
    anchorcolor=blue,
    citecolor=blue
}
\DeclareFontFamily{U}{matha}{\hyphenchar\font45}
\DeclareFontShape{U}{matha}{m}{n}{
      <5> <6> <7> <8> <9> <10> gen * matha
      <10.95> matha10 <12> <14.4> <17.28> <20.74> <24.88> matha12
      }{}
\DeclareSymbolFont{matha}{U}{matha}{m}{n}

\DeclareMathSymbol{\Lt}{3}{matha}{"CE}
\DeclareMathSymbol{\Gt}{3}{matha}{"CF}

\def\beq{\begin{equation}}
\def\eeq{\end{equation}}
\def\barr{\begin{array}}
\def\earr{\end{array}}
\def\dis{\displaystyle}

\newcommand{\be}{\begin{equation}}
\newcommand{\ee}{\end{equation}}
\newcommand{\bea}{\begin{eqnarray}}
\newcommand{\eea}{\end{eqnarray}}

\newcommand{\bi}{\begin{itemize}}
	\newcommand{\ei}{\end{itemize}}

\begin{document}

\title{Assisted baryon number violation in $4k+2$ dimensions}
\author{Akshay A}
\email{akshaymadathara1999@gmail.com}
\author{ Mathew Thomas Arun}
\email{mathewthomas@iisertvm.ac.in}
\affiliation{School of Physics, Indian Institute of Science Education and Research Thiruvananthapuram, Vithura, Kerala, 695551, India}%
\date{\today}
\begin{abstract}

Proton decay in six dimensions orbifolded on square $T^2/Z_2$ is highly suppressed at tree-level. This is because baryon number violating (BNV) operators containing only the zero mode of bulk fermions must satisfy the selection rule $\frac{3}{2} \Delta B \pm \frac{1}{2}\Delta L = 0 \ mod \ 4$. In this article, we show that the above relation does not prohibit mass dimension-6 BNV operators containing Kaluza Klein (KK) partners of the bulk fermions. Together with `spinless' adjoint scalar partner of hypercharge gauge boson (the Dark Matter candidate), these novel operators generate Dark Matter assisted proton decay at mass dimension-8. Here, with explicit examples of scalar and vector baryon number violating interactions, we discuss the importance of such $\Delta B=1 =\Delta L$ and $\Delta B=2=\Delta L$ operators and derive the limit on New Physics. 

\end{abstract}

\maketitle

\section{Introduction}

Baryon-antibaryon asymmetry in the universe has been one of the most intriguing mysteries in Nature. With the partial lifetime of proton confirmed to beyond $10^{34}$ years~\cite{Super-Kamiokande:2018apg}, it is impossible for simple baron number violating New Physics models to exist at $\mathcal{O}(1 \rm{ TeV })$. On the other hand, next order effects like neutron-antineuron ($n-\bar{n}$) oscillations~\cite{Phillips:2014fgb}, $nn \rightarrow \bar{\nu}\bar{\nu}$, Hydrogen-antiHydrogen ($H-\bar{H}$) oscillations, or double proton annihilation ($pp \rightarrow e^+e^+$), which violate baryon number by two units, have been interesting due to the much lower constraint on new physics scale. Such rare processes that violate these accidental symmetries of Standard Model (SM) have been very powerful probes to explore physics beyond Standard Model. Thus, if detected, it will be of fundamental importance in particle physics and cosmology. New experiments~\cite{Abele:2022iml} are being devised to detect these rare events. On the other hand, the lack of any conclusive observation of such events on terrestrial experiments posses indomitable challenge for New Physics models. In an effective field theory, since both scalar and vector operators lead to $\Delta B=1$ and $\Delta B=2$ processes, it is not possible to predict observable $\Delta B=2$ processes without suppressing $\Delta B=1$ with discrete symmetries or additional quantum numbers. The solution to these problems could lie in some dynamical process which suppress baryon number violating currents on Earth, but have had significant contribution to baryogengesis in the evolution of the Universe.

Though baryon and lepton numbers are accidental symmetries of Standard Model at classical level, quantum effects break them non-perturbatively \cite{tHooft:1976rip} to $U(1)_{B-L}$. There is no {\it a priori} reason for these symmetries to be preserved in beyond Standard Model scenarios. Nevertheless, to describe New Physics with minimal SM like gauge structure and representations, it is suggestive to keep the $U(1)_{B-L}$ symmetry to be intact. With the proton decay suppressed, various New Physics models~\cite{Mohapatra:1980qe, Mohapatra:1980de, Mohapatra:1996pu, Pasupathy:1982qr,Rao:1982gt, MOHAPATRA19891,Arnold:2012sd,Berezhiani:2015afa,Berezhiani:2020vbe,Arun:2022eqs,Thomas:2022hyj} can accommodate baryon number violation by two units. These processes are highly sensitive to New Physics at an intermediate energy scale $\sim \mathcal{O}(1-100 \ {\rm TeV})$. The strongest constraint on this intermediate scale arise from neutron-antineutron oscillation $\sim 500 $ TeV~\cite{Arnold:2012sd}, in four-dimensions. On the other hand, embedding the model in six-dimensions with nested warping~\cite{Thomas:2022hyj} has proven to substantially relax this constraint to $\sim 3 $ TeV. More interestingly, though $n-\bar{n}$ oscillation is usually understood to be baryon number violating by two units, with a suitably extended Higgs sector~\cite{PhysRevLett.49.7} that spontaneously break global $B-L$ symmetry, this process also violates the lepton number. With the neutrino mass $m_\nu \lesssim 10^{-1}$ eV, this model accommodates a much relaxed New Physics scales $\sim 1 $ TeV. 

The identification of 11 possible candidates with an expected background of $9.3\pm2.7$ events at Super-Kamiokande~\cite{Super-Kamiokande:2020bov}, 0.37 megaton-year exposure, and the prospect of observing the neutron-antineutron oscillation at Hyper-Kamiokande~\cite{Hyper-Kamiokande:2018ofw} and HIBEAM/NNBAR~\cite{Addazi:2020nlz} with much improved sensitivity, has reignited the interest in $\Delta B=2$ processes. Moreover, the predictions of Hydrogen-antiHydrogen oscillation and proton-proton annihilation ($pp\rightarrow e^+e^+$)~\cite{PhysRevD.18.1602} are other possible signatures of the New Physics. At the quark-level, $H-\bar{H}/pp\rightarrow e^+e^+$ is given by the dimension-12 operator, 
\begin{equation}
C_{H-\bar{H}}(uud)^2(\bar{e}^c e) \ .
\label{eq:HHquark}
\end{equation}
At the scale of the measurement ($\sim 2 $ GeV), it is convenient to construct hadron-level effective field theory as, 
\begin{eqnarray}
\mathcal{O}_{ppee} &=& \dis \frac{1}{\Lambda_{ppee}^2}\Big(\bar{p}^c \Gamma_p p\Big)\Big(\bar{e}^c \Gamma_e e \Big) \nonumber \\
\text{where} \  \ \Gamma_{p,e} &=& \dis (1,i \gamma^5,\gamma^\mu \gamma^5) \ .
\label{ppeeops}
\end{eqnarray} 
The quark-level effective operator could then be compared with the low-energy effective field theory operator made up of leptons and hadrons as,
\begin{eqnarray}
\Lambda_{ppee} (0.22 m_p)^3 = 1/\sqrt{C_{H-\bar{H}}} \ ,
\end{eqnarray}
where we have used $\langle 0| uud|p\rangle = \sqrt{2 m_p} \beta_H$, with the hadronic parameter $\beta_H = 0.014 \ {\rm GeV}^3$ determined by lattice methods~\cite{JLQCD:1999dld}. 

With large densities of atomic Hydrogen present at the interstellar medium (ISM), the search for $\Delta B=2, \ \Delta L =2$ process in the oscillation-induced diffused $\gamma-$rays survey, by Fermi LAT, constraints the upper-limit of the oscillation matrix element to $\delta = 2 \langle p e^-|H|\bar{p}e^+\rangle \lesssim 6 \times 10^{-17} s^{-1} $~\cite{Grossman:2018rdg}. On the other hand, search for $pp\rightarrow e^+e^+$ (proton annihilation rate in oxygen nuclei), at Super-Kamiokande~\cite{Super-Kamiokande:2018apg}, places a much stronger upper-bound of $\delta \lesssim 10^{-21}s^{-1}$. This limit constraints the scale of New Physics to be $\gtrsim 2$ TeV. 

In $4k+2$ dimensions, the operator in Eq.\ref{eq:HHquark} does not remain invariant under Lorentz transformations. This is because the charge conjugation operator in $4k$ and $4k+2$ dimensions behave differently. In even dimensions, it is understood that the Lorentz group is reducible and there exists a chiral projection operator. While in $4k-$dimensions, charge conjugation operator anti-commute with the chirality projection operator, in $4k+2$ dimensions they commute. Thus, it is not straight forward to realize a $4k-$dimensional model by compactifying from $4k+2$-dimensions, in the presence of currents with charge conjugate fields. In this article, we discuss the correct manner to address the baryon number violating operators in $4k+2-$dimensions, inparticular in six-dimensions. We conduct a model independent effective field theory analysis with scalar, vector and mixed operators, generated through the interactions of scalar and vector bilinear of spinor fields that transform under the full $4k+2-$dimensional Lorentz symmetry. An interesting scenario arises with operators containing KK-1 modes at the lowest order.

Note that in generic $d-$dimensions, the gauge boson has $d-2$ polarizations. After compactification, a combination of $d-4$ broken polarization in the KK-spectrum becomes the `spineless adjoint scalar field'. One such combination is ``eaten'' by the KK towers to become massive, while other combinations survive. In six dimensions, the surviving combination of the broken polarizations of hypercharge gauge boson forms the `spinless adjoint scalar', which, with the degeneracy of the KK-mode masses lifted at 1-loop~\cite{Cheng:2002iz,Ponton:2005kx}, becomes the lightest stable particle and thus the Dark Matter (DM)~\cite{Dobrescu:2007ec}. Limits from the WMAP data~\cite{WMAP:2010qai} constraints the mass of this adjoint scalar to be $\sim 2 $ TeV~\cite{Belanger:2010yx}, but it can be relaxed by allowing additional resonant annihilation and co-annihilation channels. One such possibility is to embed the model in higher dimensional space-time with warping~\cite{Arun:2018yhg}.

With the spineless adjoint scalar becoming the Dark Matter candidate, its interactions with the KK-1 fermion can influence the aforementioned operators leading to Dark Matter assisted baryon number violating currents. These operators influenced by the Dark Matter, in $4k+2-$dimensions, can predict large baryon number violation near superdense Dark Matter clumps~\cite{PhysRevD.81.103529,PhysRevD.81.103530}. This can also explain the absence of any observation yet at the terrestrial experiments. Moreover, this operator also provides an interesting annihilation channel for the Dark Matter.

In literature, four dimensional models that predict such Dark Matter influenced baryon number violation~\cite{Davoudiasl:2010am, Davoudiasl:2011fj, Blinov:2012hq, Huang:2013xfa}, are discussed usually in the context of asymmetric Dark Matter carrying a net antibaryon number which can describe both dark and baryonic matter origin through a unified phylogenesis mechanism. These antibaryonic dark matter can cause induced nucleon decay with $\sim 1 $ GeV meson in the final sate and provide a novel signature in the terrestrial nucleon decay experiments. Models with hidden MeV Dark Matter~\cite{Kile:2009nn} can also contribute to Dark matter induced processes like $\bar{f}p \to e^+ n$ and have interesting signatures at SuperKamiokande. They are constrained by Dark Matter relic density and supernova cooling, and for Majorana type Dark Matter, Super-Kamiokande strongly rules them out up to the scale $\sim 100 $ TeV. 

A minimal $4k+2-$ extra dimensional construction assumes six-dimensions, such that the six-dimensional Lorentz symmetry is broken to four dimensions by orbifolding on $T^2/Z_2$. This construction, with Standard Model like bulk fermions transforming under $SU(3)_c \times SU(2)_W \times U(1)_Y$ gauge group, boasts a rich phenomenology~\cite{Freitas:2007rh,Dobrescu:2007xf,Cacciapaglia:2011hx,Choudhury:2011jk}, provides a viable cold dark matter candidate~\cite{Dobrescu:2007ec,Cacciapaglia:2009pa}, predicts the number of chiral generations~\cite{Dobrescu:2001ae}, and can lead to small cosmological constant~\cite{Rubakov:1983bz} naturally. Upon compactification, the 6-dimensional Lorentz symmetry $SO(5,1)$ breaks to the 4-dimensional Lorentz symmetry $SO(3,1)$ and a residual $U(1)_{45}$ symmetry which generates rotation in $x_4-x_5$ plane. Thus in addition to the 4-dimensional Lorentz transformations, the fermions also transform under the $U(1)_{45}$ symmetry. This brings in additional charge to the fermions and leads to the selection rule $\frac{3}{2} \Delta B \pm \frac{1}{2} \Delta L = 0 \ mod \ 4$ for baryon and lepton number violating operator constructed only of zero mode of fermions~\cite{Appelquist:2001mj}. This selection rule suppresses the proton decay to very large orders at tree-level, and saves the model from tight constraints.

In this article, we explicitly show that this selection rule does not hold true with operators containing Kaluza Klein (KK) partners of the SM fermions. These new set of operators, although inconsequential on its own, become interesting when we include their interaction with the `spinless' adjoint scalar field. The interaction of KK-1 fermions with the `spinless' adjoint scalar field, can readily convert it to SM fermion.  Thus, here, we show that the proton can decay faster than what was discussed in~\cite{Appelquist:2001mj}, albeit in the presence of Dark Matter (`spinless' adjoint scalar field). This process then can explain the rarity of proton decay on Earth with the lack of enough Dark Matter density. 

This article is organized as follows. In the next section and its subsection, we will derive some of the relevant properties of Clifford algebra and fields $4k+2-$dimensions, particularly in six-dimensions and also discuss the field content in our model. In sec.\ref{sec:BNVops}, we discuss the possible baryon number violating interactions of scalar and vector new physics fields and resultant operators and in sec.\ref{sec:Aprotondecay}, we discuss the proton decay and assisted proton decay. The $\Delta B=2=\Delta L$, is discussed in sec.\ref{sec:B2L2} and also derive limits on these operators from various processes like $pp \to e^+e^+$, $DM+p \to DM +\bar{p}+e^++e^+$ and Dark Matter initiated Hydrogen-antiHydrogen oscillation. And we conclude our analysis in sec.\ref{sec:conclusion}.

\section{Clifford algebra in $4k+2-$dimensions}
\label{sec:clifford}
In a general d-dimensional vector space, with the basis generated by $\Gamma^M$, over the field of complex numbers, the Clifford algebra is given by~\cite{Alvarez-Gaume:1985zzv,Freund:1986ws}, 
\begin{equation}
\{\Gamma_M,\Gamma_N \} = 2 \eta_{MN} \ ,
\label{eq:clifford4k2}
\end{equation}
where $\eta_{MN} = diag(-1,+1,+1,+1,...,d-1 \text{ times})$ and $M,N = (0,1,2,3,...,d-1)$. When $d$ is even, the Clifford algebra falls apart into two simple sets, whose representations we call Weyl spinors. 

The Lorentz group generators in this geometry becomes,
\begin{equation}
\Sigma^{MN} = -\frac{i}{2}[\Gamma^M,\Gamma^N] \ .
\label{eq:4k2lorentzgens}
\end{equation}
In even dimensions, the Lorentz symmetry also supports an extra Gamma matrix that anticommutes with all the other $\Gamma^M$ as,
\begin{eqnarray}
\Gamma^{4k+3} = \alpha  \Gamma^0\Gamma^1\Gamma^2....\Gamma^{4k+1} \ ,
\label{eq:gammachiral}
\end{eqnarray}
where $\alpha = 1$ in $4k+2-$dimensions, chosen to satisfy $\Big(\Gamma^{4k+3}\Big)^2=1$.
Using this, we can define a chiral projection operator $P_{\pm} = \frac{1}{2}(1\pm \Gamma^{4k+3})$, such that every Dirac fermion ($\psi$) can be projected into two irreducible Weyl representations ($\psi_{\pm}$) by,
\begin{equation}
\psi_{\pm} = P_{\pm} \psi \ .
\label{eq:weylcond}
\end{equation}
We name the chiralities in $4k+2$-dimensions to be $+$ and $-$ to distinguish from the chiralities in $4k$-dimensions where they are called left and right. 

Moreover, for gamma matrices $\Gamma^M$, there exists similarity transformation that relates them to $-\Gamma^{M*}$. Given this transformation, we can define a charge conjugation operator that acts on the fermion field as,
\begin{equation}
\psi^c = C \psi \equiv (C \Gamma^0) \psi^* ,
\end{equation}
such that the $\psi$ and $\psi^c$ have the same Lorentz transformation, satisfying $[C\Gamma^0,\Sigma^{MN}]=0$. 
Further, the transformation of the Gamma matrix under this operator is given by,
\begin{eqnarray}
\Gamma^{M} &=& \dis - (C \Gamma^0) \Gamma^{M*} (C \Gamma^0)^{-1} \nonumber \\
&=& \dis -C (\Gamma^M)^T C^{-1} \ .
\label{eq:CCOgamma}
\end{eqnarray}
Now, from Eq.\ref{eq:gammachiral}, we see that $[C \Gamma^0, \Gamma^{4k+3}] = 0$ in $4k+2-$dimensions. Unlike in four-dimensions, since the charge conjugation operator $(C \Gamma^0)$ commutes with $\Gamma^{4k+3}$, the charge conjugate fermion representation in six-dimensions must satisfy the same Weyl condition as the original spinor field did. Due to this, the charge conjugation operator do not flip chirality in $4k+2$ dimensions.

For illustrating the arguments above, we will work with Standard Model fermions in six-dimensions and describe the relevant Lorentz symmetry properties below. 
In six-dimensions, the spin-half representation of Lorentz group is defined by six $8 \times 8$ gamma matrices that satisfy the relation in Eq.\ref{eq:clifford4k2}. In particular, we choose to work in the representation of the algebra defined by
\begin{eqnarray}
\Gamma^\mu =  \gamma^\mu \otimes \sigma^1 &,& \dis \Gamma^4 = \gamma^5 \otimes \sigma^1 \ , \Gamma^5 = \mathds{1}\otimes \sigma^2  \ .
\label{eq:gamma}
\end{eqnarray}
In the above relations, $\gamma^\mu$ denotes the four-dimensional Dirac matrices and $\gamma^5$ the chirality projection operator four-dimensions. The Lorentz algebra for the spinor field is now generated by,
\begin{eqnarray}
\Sigma_{\mu \nu} = \frac{i}{2}[\Gamma_\mu, \Gamma_\nu] &,& \dis \Sigma_{\mu 4} = \frac{i}{2}[\Gamma_\mu, \Gamma_4] \  \nonumber \\
\Sigma_{\mu 5} = \frac{i}{2}[\Gamma_\mu, \Gamma_5] &,& \dis \Sigma_{4 5} = \frac{i}{2}[\Gamma_4, \Gamma_5] \ ,
\label{eq:6dLorentzgen}
\end{eqnarray}
with spinors transforming as $\Psi \rightarrow e^{\frac{i}{4} \Sigma_{MN} \theta^{MN}} \Psi$. Like discussed previously, Lorentz group in six-dimension admits irreducible chiral representations $\Psi_{\pm} = \frac{1}{2}(1 \pm \Gamma^7) \Psi$, where the chiral projection operator is given by,
\begin{equation}
\Gamma^7 =  \Gamma^0\Gamma^1\Gamma^2\Gamma^3\Gamma^4\Gamma^5 = \mathds{1}\otimes \sigma^3 \ .
\label{eq:gamma}
\end{equation}

Using Eq.\ref{eq:CCOgamma}, along with the gamma matrices given in Eq.\ref{eq:gamma}, the charge conjugation operator $C$ can be seen to anti-commute with $\Gamma^0,\Gamma^2, \Gamma^4$ and commute with $\Gamma^1,\Gamma^3,\Gamma^5$. Therefor the charge conjugation operator is given by,
\begin{eqnarray}
C &=& \dis i\Gamma^4\Gamma^2\Gamma^0 \nonumber \\
&=& \dis \gamma^5\gamma^2\gamma^0 \otimes \sigma^1 \ .
\label{eq:chargeconjop}
\end{eqnarray}
\subsection{Standard Model fermions in six-dimensions}
Lets now consider bulk Standard Model fermions in six-dimensions that transform under $SU(3)_c\times SU(2)_W \times U(1)_Y$ gauge group. These fermions are denoted by $\mathcal{Q}_+$, $\mathcal{U}_-$, $\mathcal{D}_-$ for quarks and $L_{+}$, $\mathcal{E}_{-}$, $\mathcal{N}_{-}$ for leptons, where $\pm$ are chiralities defined by the chirality projection operator defined as $P_{\pm} = \frac{1}{2}\Big(1 \pm \Gamma^7 \Big)$. These fields are set to satisfy the gauge quantum numbers given in Table.\ref{tab:6dSM}.
\begin{table}
\centering
\begin{tabular}{|c|c|c|c|}
\hline 
fermions & $SU(3)_c$ & $SU(2)_W$ & $U(1)_Y$ \\
[1ex] 
\hline
$\mathcal{Q}_+ (x^M)$ & 3 & 2 & 1/3 \\[1ex]
$\mathcal{U}_-(x^M)$ & 3 & 1 & 4/3 \\[1ex]
$\mathcal{D}_-(x^M)$ & 3 & 1 & -2/3 \\[1ex]
$L_{+}(x^M)$ & 1 & 2 & -1 \\[1ex]
$\mathcal{E}_{-}(x^M)$ & 1 & 1 & -2\\[1ex]
$\mathcal{N}_{-}(x^M)$ & 1 & 1 & 0\\[1ex]
\hline
\end{tabular}

\caption{Six-dimensional Standard Model fermions and their charges under $SU(3)_c\times SU(2)_W \times U(1)_Y$ gauge group.}
\label{tab:6dSM}
\end{table}

Unlike in 4 dimensions, in six dimensions the gauge anomalies are give by box diagrams.
The afore mentioned fermion chiralities and gauge charges are assigned such that the irreducible $[SU(3)_c]^3 U(1)_Y$ and mixed gauge-gravitational anomalies vanish exactly. But, the non-vanishing reducible anomalies, $[SU(2)_W]^4$, $[SU(2)_W]^2[SU(3)_c]^2$,  $[SU(2)_W]^2[U(1)]_Y^2$ and $[SU(3)_c]^2[U(1)]_Y^2$ are cancelled via Green-Schwarz mechanism~\cite{Erler:1993zy}.

On compactifying the six-dimensional geometry on a torus $T^2$, the Lorentz generators in Eq.\ref{eq:6dLorentzgen} breaks to $\Sigma_{\mu \nu}$ and $\Sigma_{45}$. Note that, $\Sigma_{45}$ generates rotation in the $x_4-x_5$ plane. Hence, along with the four-dimensional Lorentz transformation, $\Psi_{\pm} \rightarrow  e^{\frac{i}{4} \Sigma_{\mu \nu} \theta^{\mu \nu}} \Psi_{\pm}$, the fermions also transform under $\Sigma_{45}$ as $\Psi_{\pm} \rightarrow e^{\frac{i}{4} \Sigma_{45} \theta^{45}} \Psi_{\pm}$. This residual $U(1)_{45} = e^{i \Sigma^{45} \theta_{45}}$ symmetry, where $\theta_{45}$ is an arbitrary rotation in $(x_4,x_5)$ plane, is broken to its discrete subgroups upon orbifolding. The $T^2/Z_2$ orbifold on a rectangle, in general, is now invariant under a rotation through $\pi$, thus preserving the $Z_2$ subgroup of $U(1)_{45}$. Whereas, a square $T^2/Z_2$ posses a $Z_4$ symmetry since it is invariant under $\pi/2$ rotations. 

Orbifolding six-dimensions on $T^2/Z_2$ breaks the six-dimensional fermion $\Psi_{\pm}(x^{M})$ to its Fourier mode $\Psi_{\pm}(x^M) = \sum_{n} \psi_{\pm l}^{n}(x^{\mu}) \chi_l^{n}(x^4,x^5) +  \psi_{\pm r}^{n}(x^{\mu}) \chi_r^{n}(x^4,x^5)$, where $l$ and $r$ are four-dimensional chiralities are given by $\psi_{\pm l} = P_L \Psi_{\pm} = \frac{1}{2}(1+\gamma^5)\Psi_{\pm}$ and $\psi_{\pm r} = P_R \Psi_{\pm} = \frac{1}{2}(1 - \gamma^5)\Psi_{\pm}$, where $\gamma_5 = i \gamma_0\gamma_1\gamma_2\gamma_3$. Since the residual $Z_4$ subgroup of $U(1)_{45}$ is preserved under orbifolding, the fermions $\psi_{\pm l}$  gets a charge $\pm 1/2$ and $\psi_{\pm r}$ gets charge $\mp 1/2$ under this symmetry~\cite{Appelquist:2001mj}. Thus, all the operators originating from six-dimensional geometry are bound to preserve this quantum charge.\\[-5.1ex]

To understand what this means for six-dimensions with bulk Standard Model, lets consider their Kaluza Klein decomposition,
 \begin{eqnarray}
 \mathcal{Q}_{+} (x^\mu,x_a) & = &  \frac{\sqrt{2}}{(2\pi
R)}\left\{q^{(0,0)}_{+l}(x^\mu) + \sqrt{2} \sum_{m,n} \right.
\left[P_L \mathcal{Q}_{+l}^{(m,n)}(x^\mu) \, \cos\left( \frac{1}{R}(m x_4 +
n x_5) \right) \right. \nonumber \\ [0.5em] && + \left.\left.
P_R \mathcal{Q}_{+r}^{(m,n)}(x^\mu) \, \sin\left( \frac{1}{R}(m x_4
+ n x_5) \right)\right] \right\} ~,
 \nonumber \\ [1em]
\mathcal{U}_{-} (x^\mu,x_a) & = & \frac{\sqrt{2}}{(2\pi R)}
\left\{ u^{(0,0)}_{-r}(x^\mu) + \sqrt{2} \sum_{m,n}\right.  \left[ P_R \mathcal{U}_{-r}^{(m,n)}(x^\mu) \, \cos\left( \frac{1}{R}(m x_4 +n x_5) \right) \right. \nonumber \\ [0.5em] && + \left.\left.
P_L \mathcal{U}_{-l}^{(m,n)}(x^\mu) \, \sin\left( \frac{1}{R}(m x_4
+ n x_5) \right) \right] \right\} \nonumber\\
\mathcal{D}_{-} (x^\mu,x_a) & = & \frac{\sqrt{2}}{(2\pi R)}
\left\{ d^{(0,0)}_{-r}(x^\mu) + \sqrt{2} \sum_{m,n}\right.  \left[ P_R \mathcal{D}_{-r}^{(m,n)}(x^\mu) \, \cos\left( \frac{1}{R}(m x_4 +n x_5) \right) \right. \nonumber \\ [0.5em] && + \left.\left.
P_L \mathcal{D}_{-l}^{(m,n)}(x^\mu) \, \sin\left( \frac{1}{R}(m x_4
+ n x_5) \right) \right] \right\} \nonumber\\
L_{+} (x^\mu,x_a) & = &  \frac{\sqrt{2}}{(2\pi
R)}\left\{\ell^{(0,0)}_{+ l}(x^\mu) + \sqrt{2} \sum_{m,n} \right.
\left[P_L L_{+ l}^{(m,n)}(x^\mu) \, \cos\left( \frac{1}{R}(m x_4 +
n x_5) \right) \right. \nonumber \\ [0.5em] && + \left.\left.
P_R L_{+ r}^{(m,n)}(x^\mu) \, \sin\left( \frac{1}{R}(m x_4
+ n x_5) \right)\right] \right\} ~,\nonumber\\
\mathcal{E}_{-} (x^\mu,x_a) & = & \frac{\sqrt{2}}{(2\pi R)}
\left\{ e^{(0,0)}_{- r}(x^\mu) + \sqrt{2} \sum_{m,n}\right.  \left[ P_R \mathcal{E}_{- r}^{(m,n)}(x^\mu) \, \cos\left( \frac{1}{R}(m x_4 +n x_5) \right) \right. \nonumber \\ [0.5em] && + \left.\left.
P_L \mathcal{E}_{- l}^{(m,n)}(x^\mu) \, \sin\left( \frac{1}{R}(m x_4
+ n x_5) \right) \right] \right\} ~,\nonumber\\
\mathcal{N}_{-} (x^\mu,x_a) & = & \frac{\sqrt{2}}{(2\pi R)}
\left\{ n^{(0,0)}_{- r}(x^\mu) + \sqrt{2} \sum_{m,n}\right.  \left[ P_R \mathcal{N}_{- r}^{(m,n)}(x^\mu) \, \cos\left( \frac{1}{R}(m x_4 +n x_5) \right) \right. \nonumber \\ [0.5em] && + \left.\left.
P_L \mathcal{N}_{- l}^{(m,n)}(x^\mu) \, \sin\left( \frac{1}{R}(m x_4
+ n x_5) \right) \right] \right\}
\label{eq:ffermion}
 \end{eqnarray}
where $q^{(0,0)}_{+l}$, $u^{(0,0)}_{-r}$ and $d^{(0,0)}_{-r}$ are the zero modes and are identified with the four-dimensional Standard Model quarks. Similarly, $\ell^{(0,0)}_{+ l}$, $e^{(0,0)}_{- r}$ and $n^{(0,0)}_{- r}$ are identified with the Standard Model leptons and the right handed neutrino. The rest of the states are Kaluza Klein partners of the Standard Model fermions and carry the same gauge quantum charge. As mentioned before, the fermions are also charged under the residual $U(1)_{45}$ symmetry. 
Since $\Sigma_{45} = \gamma^5\otimes \sigma^3$, from Eq.\ref{eq:6dLorentzgen}, left and right handed partners of the same fermion carry opposite charge. The full set of charges four-dimensional fermion fields carry are given in Table.\ref{tab:SMqc}.  
\begin{table}[!h]
\centering
\begin{tabular}{|c|c|c c c|c|}
\hline 

6-d fermions & 4-d fermions & $SU(3)_c$ & $SU(2)_W$ & $U(1)_Y$ & $U(1)_{45}$\\
[1ex] 
\hline
\multirow{2}{0.07\columnwidth}{$\mathcal{Q}_+ (x^M)$} & $q^{(0,0)}_{+l}(x^\mu)$, $\mathcal{Q}^{(m,n)}_{+l}(x^\mu)$ & \multirow{2}{0.01\columnwidth}{3} & \multirow{2}{0.01\columnwidth}{2} & \multirow{2}{0.04\columnwidth}{1/3} & 1/2 \\ [1ex]
& $\mathcal{Q}^{(m,n)}_{+r}(x^\mu)$ &  &  &  & - 1/2 \\[1ex] \hline
\multirow{2}{0.07\columnwidth}{$\mathcal{U}_- (x^M)$} & $u^{(0,0)}_{-r}(x^\mu)$, $\mathcal{U}^{(m,n)}_{-r}(x^\mu)$ & \multirow{2}{0.01\columnwidth}{3} & \multirow{2}{0.01\columnwidth}{1} & \multirow{2}{0.04\columnwidth}{4/3} & 1/2\\[1ex]
& $\mathcal{U}^{(m,n)}_{-l}(x^\mu)$  & & & & -1/2\\[1ex] \hline
\multirow{2}{0.07\columnwidth}{$\mathcal{D}_- (x^M)$} & $d^{(0,0)}_{-r}(x^\mu)$,  $\mathcal{D}^{(m,n)}_{-r}(x^\mu)$ & \multirow{2}{0.01\columnwidth}{3} & \multirow{2}{0.01\columnwidth}{1} & \multirow{2}{0.06\columnwidth}{-2/3} & 1/2\\[1ex]
& $\mathcal{D}^{(m,n)}_{-l}(x^\mu)$ & & & & -1/2\\[1ex] \hline
\multirow{2}{0.07\columnwidth}{$L_+ (x^M)$} &  $\ell^{(0,0)}_{+l}(x^\mu)$, $L^{(m,n)}_{+l}(x^\mu)$ & \multirow{2}{0.01\columnwidth}{1} & \multirow{2}{0.01\columnwidth}{2} & \multirow{2}{0.05\columnwidth}{-1} & 1/2\\[1ex]
& $L^{(m,n)}_{+r}(x^\mu)$  & & & & -1/2\\[1ex] \hline
\multirow{2}{0.07\columnwidth}{$\mathcal{E}_- (x^M)$} &  $e^{(0,0)}_{-r}(x^\mu)$,  $\mathcal{E}^{(m,n)}_{-r}(x^\mu)$ & \multirow{2}{0.01\columnwidth}{1} & \multirow{2}{0.01\columnwidth}{1} & \multirow{2}{0.04\columnwidth}{-2} & 1/2\\[1ex]
& $\mathcal{E}^{(m,n)}_{-l}(x^\mu)$ & & & & -1/2\\[1ex] \hline
\multirow{2}{0.07\columnwidth}{$\mathcal{N}_- (x^M)$} &  $n^{(0,0)}_{-r}(x^\mu)$, $\mathcal{N}^{(m,n)}_{-r}(x^\mu)$ & \multirow{2}{0.01\columnwidth}{1} & \multirow{2}{0.01\columnwidth}{1} & \multirow{2}{0.03\columnwidth}{0} & 1/2\\[1ex]
& $\mathcal{N}^{(m,n)}_{-l}(x^\mu)$ & & & & -1/2\\[1ex] 
\hline
\end{tabular}
\caption{Resultant charges of four-dimensional fermions after breaking the six-dimensional Lorentz symmetry, $SO(5,1)$, to $SO(3,1) \times U(1)_{45}$.}
\label{tab:SMqc}
\end{table}

\subsection{New Physics Scalar and Vector bosons in six-dimensions}
\label{sec:scalarandvectorNP}
The Higgs field and $SU(3)_c \times SU(2)_W \times U(1)_Y$ gauge bosons are assumed to propagate in the bulk of six-dimensions. Upon compactification and orbifolding, they break to KK state. For brevity, we will not discuss the Standard Model scalar and vector fields, but refer the reader to \cite{Appelquist:2000nn}. On the other hand, since `spinless' adjoint scalar partner of the hypercharge gauge boson is of importance, we will discuss the abelian gauge theory in six-dimensions in Appendix.\ref{section:A3}.

Let us now consider the new physics scalar and vector bosons that mediate the baryon number violating currents. Their charges under Standard Model gauge group are given in Table.\ref{tab:vectorint}. We will assume that unlike Standard Model fields these new physics bosons satisfy the Dirichlet boundary condition at orbifold fixed points. The reason for this will be clear when we write their interaction terms. 
{\renewcommand{\arraystretch}{1.4}\begin{table}[t!]
\begin{center}
    \begin{tabular}{| c || c | c | c |}
    \hline
       \ \ Field  \ \  & \raisebox{0ex}[0pt]{$ \ \ {\rm SU}(3)_c \ \ $} & \raisebox{0ex}[0pt]{$ \ \  {\rm SU}(2)_L \ \ $} & \raisebox{0ex}[0pt]{$ \ \ \ {\rm U}(1)_Y \ \ \ $}  \ \ \ 
       \\ \hline\hline   
    \ \ $\Phi^{\dagger}   $ \ \  & $ 3 $ & $2$ & $ 5/3 $\\ \hline
          \ \ $A_M   $ \ \  & $ \bar{6} $ & $3$ & $ 2/3 $\\ \hline
      \ \  $V_M^* $ \ \  & $3 $ & $1,3$ & $ 2/3 $ \\ \hline
      \ \ $U_M   $ \ \  & $ 3 $ & $1$ & $ 2/3 $\\ \hline
\end{tabular}
\end{center}\vspace{-2mm}
\caption{Six-dimensional scalar and vector color representations in addition to Standard Model fields.}\vspace{1mm}
\label{tab:vectorint}
\end{table}}

To that end, lets begin with the Lagrangian of a complex scalar boson ($\Phi(x_\mu,x_4,x_5)$) in six-dimensions given by,
\bea
\mathcal{L}_{S} = \Big(D_M \Phi\Big)^\dagger \Big(D^M\Phi\Big) - m_{\Phi}^2 \Phi^\dagger \Phi \ ,
\eea
where $D_M$ is the covariant derivate. Orbifolding the geometry and applying Dirichlet boundary condition, $\Phi$ can be expanded in its Fourier modes as,
\bea
\Phi(x_\mu,x_4,x_5) = \sum_{m,n=0} \Phi^{(m,n)}(x_\mu) \, \sin\left(\frac{m x_4}{R}+\frac{n x_5}{R} \right) \ .
\label{eq:scalarKKmodes}
\eea
With this, after integrating out the extra dimensions, the four-dimensional Lagrangian density becomes,
\bea
\mathcal{L}_{S 4D} = \int dx_4 dx_5 \mathcal{L}_{S} = \Big(D_\mu \Phi^{(m,n)}\Big)^\dagger \Big(D^\mu\Phi^{(m,n)} \Big) - \Big(\frac{1}{R^2}(m^2 + n^2) + m_{\Phi}^2  \Big)\Phi^{(m,n)\dagger} \Phi^{(m,n)} \ .
\eea
Note that the lightest mode in the above Lagrangian have mass $\Big(\frac{1}{R^2} + m_{\Phi}^2\Big)^{1/2}$. 

Similarly, the Lagrangian density of the vector boson ($A_M$) in six-dimensions is given by,
\begin{eqnarray}
\mathcal{L}_A &=& \dis -\frac{1}{4}F_{MN}F^{MN} -m_A^2 A_M A^M\nonumber \\
 &=& \dis-\frac{1}{4}F_{\mu \nu}F^{\mu \nu} - \frac{1}{2}F_{\mu 4} F^{\mu 4}- \frac{1}{2}F_{\mu 5} F^{\mu 5} - \frac{1}{2}F_{45}F^{45} -m_A^2 A_\mu A^\mu - m_A^2 A_4A^4 -m_A^2 A_5A^5 \ , \nonumber \\
\label{eq:lagrangian6D_exp}
\end{eqnarray}
where $F_{\mu 4} = \partial_\mu A_4 - \partial_4 A_\mu$, $F_{\mu 5} = \partial_\mu A_5 - \partial_5 A_\mu$, and $F_{45} = \partial_4 A_5 - \partial_5 A_4$. For simplicity we have considered Abelian fields here, but the derivation can be extended to non-Abelian fields as well.

Upon orbifolding, we again consider Dirichlet boundary condition for the $A_\mu$ component of the vector field , while $A_4$ and $A_5$ are set to satisfy Neumann boundary condition. The components of the 6D vector field, that satisfy the above boundary condition in the orbifolded geometry, now can be Fourier expanded as,
\begin{eqnarray}
A_\mu(x_\mu,x_4,x_5) = \sum_{m,n \neq 0} A^{(m,n)}_\mu(x_\mu) \, \sin\left(\frac{m x_4}{R}+\frac{n x_5}{R} \right) \ , \nonumber \\
A_4(x_\mu,x_4,x_5) = \sum_{m,n = 0} A^{(m,n)}_4(x_\mu) \, \cos\left(\frac{m x_4}{R}+\frac{n x_5}{R} \right) \ , \nonumber \\
A_5(x_\mu,x_4,x_5) = \sum_{m,n = 0} A^{(m,n)}_5(x_\mu) \, \cos\left(\frac{m x_4}{R}+\frac{n x_5}{R} \right) \ , 
\label{eq:KKvectorboson}
\end{eqnarray}

The Lagrangian density in Eq.\ref{eq:lagrangian6D_exp}, in generalized $R_\zeta$ gauge-fixing, after integrating over the $x_4$ and $x_5$ directions, become,
\begin{eqnarray}
\mathcal{L}_A &=& \dis  -\frac{1}{4}F_{\mu \nu}F^{\mu \nu} - \frac{1}{2}F_{\mu 4} F^{\mu 4}- \frac{1}{2}F_{\mu 5} F^{\mu 5} - \frac{1}{2}F_{45}F^{45} -m_A^2 A_\mu A^\mu - m_A^2 A_4A^4 -m_A^2 A_5A^5 \nonumber \\
 && \dis -\frac{1}{2 \zeta} (\partial_\mu A^\mu + \zeta(\partial_4 A^4 + \partial_5 A_5))^2\nonumber \\
&=& \dis \sum_{m,n \neq 0}-\frac{1}{4}F^{(m,n)}_{\mu \nu}F^{(m,n)\mu \nu} - \frac{1}{2} \Big((\frac{m}{R})^2+(\frac{n}{R})^2 \Big)A_\mu^{(m,n)}A^{(m,n)\mu} - m_A^2 A^{(m,n)}_\mu A^{(m,n)\mu} \nonumber \\
&& \dis -\frac{1}{2\zeta} (\partial_\mu A^{(m,n)\mu})^2  \nonumber \\
&& \dis +\sum_{m,n = 0}-\frac{1}{2}\Big(\partial_\mu A_4^{(m,n)}\Big)^2 - \frac{1}{2}\Big(\partial_\mu A_5^{(m,n)}\Big)^2 \nonumber \\
&&- \frac{1}{2}\Big(\frac{n}{R} A^{(m,n)}_4 - \frac{m}{R} A^{(m,n)}_5\Big)^2 -\frac{\zeta}{2}\Big(\frac{m}{R}A^{(m,n)}_4 + \frac{n}{R}A^{(m,n)}_5  \Big)^2 -m_A^2 A_4^{(m,n)2} -m_A^2 A_5^{(m,n)2} .\nonumber \\
\label{eq:lagrangian}
\end{eqnarray}
Since in this article, we are only interested in the first few heavy modes of the field, the dynamics of zeroth and first KK sates, represented by $(m=0,n=0)$ \& $(m=1,n=0)$ is given by
\begin{eqnarray}
\mathcal{L}_A &=& \sum_{0<m \le 1} -\frac{1}{4}F^{(m,0)}_{\mu \nu}F^{(m,0)\mu \nu} - \frac{1}{2} \Big(\frac{m}{R}\Big)^2A_\mu^{(m,0)}A^{(m,0)\mu} - m_A^2 A^{(m,0)}_\mu A^{(m,0)\mu} -\frac{1}{2\zeta} (\partial_\mu A^{(m,0)\mu})^2 \nonumber \\
&& \sum_{0 \le m \le 1}- \frac{1}{2}\Big(\partial_\mu A_4^{(n,0)}\Big)^2 - \tilde{m}_{\zeta A m}  A^{(m,0)2}_4  - \frac{1}{2}\Big(\partial_\mu A_5^{(m,0)}\Big)^2 
 - \tilde{m}_{A m}^2 A^{(m,0)2}_5 , 
\label{eq:firstfewKKvector}
\end{eqnarray}
where $\tilde{m}_{\zeta A m} = \zeta \Big(\frac{m}{R}\Big)^2 + m_A^2$ and $\tilde{m}_{A m} = \Big(\frac{m}{R}\Big)^2 + m_A^2$.

\subsection{Baryon number violating interactions of the New Physics scalar and vector boson}

In this subsection, we discuss the baryon number violating interactions of the New Physics described above. To start with, let us consider the interaction of the scalar boson. The interaction terms in the Lagrangian density is given by,
\bea
\Big(\mathcal{L}_{int}\Big)_{\Phi} &=& \dis \overline{\mathcal{Q}}^c_+ \mathcal{U}_- \Phi + \overline{\mathcal{E}}^c_- \mathcal{Q}_+ \Phi^\dagger \nonumber \\
& = & \dis \mathcal{Q}^T_+ C \mathcal{U}_- \Phi(x_\mu,x_4,x_5) + \mathcal{E}^T_- C \mathcal{Q}_+ \Phi^\dagger(x_\mu,x_4,x_5) \ ,
\eea
where $\overline{\mathcal{Q}}^c_+ = \overline{C \bar{\mathcal{Q}}_+^T}=(C \bar{\mathcal{Q}}_+^T)^{\dagger}\Gamma^0= \bar{\mathcal{Q}}_+^* C^{\dagger} \Gamma^0 = \mathcal{Q}_+^T \Gamma^0 C^{\dagger} \Gamma^0 = \mathcal{Q}_+^T C$. Here, I have used the property, $\Gamma^{M \dagger} = \Gamma^0 \Gamma^M \Gamma^0$ and charge conjugation operator $C$ is defined in Eq.\ref{eq:chargeconjop}. 

Using the Fourier decompositions given in Eq.\ref{eq:ffermion} and Eq.\ref{eq:scalarKKmodes}, after integrating out the $x_4$ and $x_5$ directions, the above interactions contains,
\bea
\int dx_4 dx_5 \Big(\mathcal{L}_{int}\Big)_{\Phi} &\supset& \dis \sum_{n,m \neq 0}q^{(0,0)T}_{+l}\gamma^2\gamma^0  \mathcal{U}^{(m,n)}_{-l} \Phi^{(m,n)}(x_\mu) \nonumber \\
&& \dis + \sum_{n,m \neq 0}\mathcal{E}^{(m,n)T}_{-l}\gamma^2\gamma^0  q^{(0,0)}_{+l} \Phi^{(m,n)\dagger}(x_\mu)
\label{eq:scalarint}
\eea
The rest of the terms are not of interest to us. 

Similarly,  for the vector bosons, the six-dimensional interaction is given by,
\begin{eqnarray}
\Big(\mathcal{L}_{int}\Big)_{U} &=& \dis \overline{\mathcal{U}}^c_- \Gamma^M \mathcal{D}_- U_M  = \mathcal{U}^T_- C \Gamma^M \mathcal{D}_- U_M  \nonumber \\
&=& \dis \mathcal{U}^T_- C  \Gamma^\mu \mathcal{D}_- U_\mu(x_\mu,x_4,x_5) + \mathcal{U}^T_- C  \Gamma^4 \mathcal{D}_- U_4(x_\mu,x_4,x_5) + \mathcal{U}^T_- C  \Gamma^5 \mathcal{D}_- U_5(x_\mu,x_4,x_5) \ . \nonumber \\
\label{eq:6DBNVinteraction}
\end{eqnarray}
Here, we have used the interaction terms of $U_M(x^N)$ as an example.

Using the KK decomposition of the fermions and vector boson given in Eq.\ref{eq:ffermion} and Eq.\ref{eq:KKvectorboson}, after integrating out the $x_4$ and $x_5$ directions, the non-vanishing contributions of the fermion KK states contains,
\bea
\int dx_4 dx_5 \Big(\mathcal{L}_{int}\Big)_{U} & \supset & \dis  -\sum_{m,n \neq 0} u^{(0,0)T}_{-r} \gamma^2\gamma^0 \gamma^\mu \mathcal{D}_{-l}^{(m,n)} U_\mu^{(m,n)}(x_\mu) \\
 & \supset & \dis  -u^{(0,0)T}_{-r} \gamma^2\gamma^0 \gamma^\mu \mathcal{D}_{-l}^{(1,0)} U_\mu^{(1,0)}(x_\mu) \ .
\eea
Since we are only interested in the lightest KK partner of the vector boson, most dominant contribution to the baryon number violating operator arises from the interaction term,

\bea
\int dx_4 dx_5 \Big(\mathcal{L}_{int}\Big)_{U} & \supset & \dis - u^{(0,0)T}_{-r} \gamma^2\gamma^0 \gamma^\mu \mathcal{D}_{-l}^{(1,0)} U_\mu^{(1,0)}(x_\mu) \ .
\label{eq:unitaryinteractionU}
\eea

This analysis can be generalised to other vector fields given in Table.\ref{tab:vectorint} and we get, 
\bea
\int dx_4 dx_5 \Big( \mathcal{L}_{int}\Big)_{A} & \supset & \dis - q^{(0,0)T}_{+l} \gamma^2\gamma^0 \gamma^\mu \mathcal{Q}_{+r}^{(1,0)} A_\mu^{(1,0)}(x_\mu)  \ .
\label{eq:unitaryinteractionA}
\eea
\bea
\int dx_4 dx_5 \Big( \mathcal{L}_{int} \Big)_{V^*} & \supset & \dis - q^{(0,0)T}_{+l} \gamma^2\gamma^0 \gamma^\mu L_{+r}^{(1,0)} V_\mu^{(1,0)*}(x_\mu) \ .
\label{eq:unitaryinteractionVstar}
\eea
\bea
\int dx_4 dx_5 \Big( \mathcal{L}_{int} \Big)_{U^*} & \supset & \dis -u^{(0,0)T}_{-r} \gamma^2\gamma^0 \gamma^\mu \mathcal{E}_{-l}^{(1,0)} U_\mu^{(1,0)*}(x_\mu)  \ .
\label{eq:unitaryinteractionUstar}
\eea

Integrating out these heavy bosons generate the operators that violate baryon number violation. But these operators contain KK-1 modes of the Standard Model fermions. As an example, from Eq.\ref{eq:scalarint}, integrating out $\Phi^{(1,0)}(x_\mu)$ we obtain the operator, 
\[
(q^{T(0,0)}_{+l}\gamma^0\gamma^2U_{-l}^{(1,0)})(\mathcal{E}_{-l}^{T(1,0)}\gamma^0\gamma^2q^{(0,0)}_{+l}) \ .
\]
 And from Eq.\ref{eq:unitaryinteractionU} and Eq.\ref{eq:unitaryinteractionUstar}, on integrating out $U_\mu^{(1,0)}(x_\mu)$, we get the corresponding vector operator. One important observation is that, there are no direct proton decay operator at tree level.

But, before we conclude this section, it is important that we discuss the interactions of the `spinless' adjoint scalar.

\subsection{Interaction of `spinless' adjoint scalar ($V_B^{(1,0)}$) with fermions}
\label{section:A4}
For illustration, let us consider the interaction term in the 6-dimensional Lagrangian density for the `spinless' adjoint scalar with quark field. Details required for this subsections is given in Appendix.\ref{section:A3}. 

The kinetic term of the quarks are given by,
\begin{equation}
    \mathcal{L}_f=\mathcal{\bar{Q}}_+(x^M)\Gamma^M D_M\mathcal{Q}_+(x^M) +\mathcal{\bar{U}}_-(x^M)\Gamma^M D_M \mathcal{U}_-(x^M) \ ,
\end{equation}
where $\Gamma^M$ are the 6-dimensional gamma matrices and the covariant derivatives are defined as $D_M\mathcal{Q}_+=(\partial_M-\frac{ig}{2}\tau^i W^i_M-i\frac{g_Y}{2}y_+ B_M)\mathcal{Q}_+$, \& $D_M\mathcal{U}_-=(\partial_M-\frac{ig_Y}{2}y_- B_M)\mathcal{U}_-$. The hypercharge interaction term in the above Lagrangian is given by,
\begin{equation}
    \mathcal{L}_I=\frac{ig_Y}{2}(y_+\mathcal{\bar{Q}}_+ \Gamma^M B_M\mathcal{Q}_+ + y_-\mathcal{\bar{U}}_- \Gamma^M B_M\mathcal{U}_-) \ .
\end{equation}
Using the KK expansions given in Eq.\ref{eq:ffermion} and Eq.\ref{eq:SMgaugeKK} for quarks and Standard Model gauge bosons, and integrating out the extra-dimensions, the above interaction term contain the term,
\begin{eqnarray}
    \mathcal{L}_I &\supset & \dis \frac{ig_Y}{2} \Big( B_4^{(1,0)}(y_{q} \bar{\mathcal{Q}}^{(1,0)}_{+r} \gamma^5 q_{+l}+y_{u} \bar{\mathcal{U}}^{(1,0)}_{-l} \gamma^5  u_{-r}) \nonumber \\
& + & \dis B_5^{(1,0)}(y_{q} \bar{\mathcal{Q}}^{(1,0)}_{+r} q_{+l} - y_{u} \bar{\mathcal{U}}^{(1,0)}_{-l}  u_{-r}) \ .
\end{eqnarray}
On diagonalising using Eq.\ref{adj}, for $(m=1,n=0)$, the term becomes,
\begin{eqnarray}
    \mathcal{L}_I &\supset & \dis \frac{ig_Y}{2} \Big( V_1^{(1,0)}(y_{q} \bar{\mathcal{Q}}^{(1,0)}_{+r} q_{+l} - y_{u} \bar{\mathcal{U}}^{(1,0)}_{-l} u_{-r}) \nonumber \\
& + & \dis V_2^{(1,0)}(y_{q} \bar{\mathcal{Q}}^{(1,0)}_{+r} q_{+l} - y_{u} \bar{\mathcal{U}}^{(1,0)}_{-l} u_{-r}) \ .
\end{eqnarray}
Since $V_1^{(1,0)}$ field is non dynamical, in unitary gauge, the interaction of the `spinless' adjoint is given by,
\begin{eqnarray}
    \mathcal{L}_I &\supset & \dis V_2^{(1,0)}(y_{q} \bar{\mathcal{Q}}^{(1,0)}_{+r} q_{+l} - y_{u} \bar{\mathcal{U}}^{(1,0)}_{-l}  u_{-r}) \ .
\end{eqnarray}
Upon identifying the $V_2^{(1,0)}$ with the Dark Matter and re-naiming it to $V_B^{(1,0)}$, we see that the non-vanishing terms in the interaction between the $V_B^{(1,0)}$ and quarks take the form,
\begin{equation}
    \mathcal{L}_I=\frac{ig_Y}{2} V_B^{(1,0)}(y_{q} \bar{\mathcal{Q}}^{(1,0)}_{+r} q_{+l} - y_{u} \bar{\mathcal{U}}^{(1,0)}_{-l} u_{-r})
    \label{eq:DMqint}
\end{equation}
Similarly, for leptons, the interaction becomes,
\begin{equation}
    \mathcal{L}_I=\frac{ig_Y}{2} V_B^{(1,0)}(y_{\ell} \bar{L}^{(1,0)}_{+r} \ell_{+l} - y_{e} \bar{\mathcal{E}}^{(1,0)}_{-l} e_{-r})
    \label{eq:DMeint}
\end{equation}
These terms shown here satisfy the quantum charge of $U(1)_{45}$ given in Table.\ref{tab:SMqc}.

\section{Baryon number violating operators}
\label{sec:BNVops}

In this section, we discuss the baryon number violating operators generated by integrating out the new physics scalar and vector bosons discussed in Sec.\ref{sec:scalarandvectorNP}. Though we work with six-dimensions to illustrate our arguments, the operators and results can be generalised to any $4k+2$-dimensions. 
The operators to be discussed here are a consequence of the interactions discussed in Eq.\ref{eq:scalarint} and Eq.\ref{eq:unitaryinteractionU} - Eq.\ref{eq:unitaryinteractionUstar}.

Before, going to the main discussion of this article, for completeness, let us first discuss why there are no tree-level proton decay operators~\cite{Appelquist:2001mj}, containing only the zero modes of fermions. Since each quark carries $1/3$ baryon number, the operator mediating baryon number violation contains $3\Delta B$ quarks. From Table.\ref{tab:SMqc}, note the zero mode of quarks carry an additional $1/2$ charge under $U(1)_{45}$. Hence, a baryon number violating operator, under this residual symmetry, has charge $\frac{3}{2}\Delta B$. Assuming that the lepton number is also violated in the process, the operator carries $\frac{3}{2}\Delta B \pm \frac{1}{2} \Delta L$ charge under $U(1)_{45}$. Since the $T^2/Z_2$ orbifold breaks the $U(1)_{45}$ symmetry down to a discrete subgroup, the operators constructed only with the zero modes now must satisfy the selection rule,
\begin{equation}
  \frac{3}{2} \Delta B \pm \frac{1}{2}\Delta L = 0 \ mod \ 4 \ .
  \label{eq:squaresr}
\end{equation}
This makes sure that baryon number violating operators appear only at dimension 15~\cite{Appelquist:2001mj}. 

On the other hand, the quantum charges in Table\ref{tab:SMqc} clearly show that this relation do not hold true once KK modes are introduced in the external legs. Below, we will derive the baryon number violating operators with KK modes and we will show that such novel operators become relevant and interesting on introducing interactions with `spinless' adjoint scalar partner of hypercharge gauge boson. 

Using the interaction terms given in Eq.\ref{eq:scalarint}, Eq.\ref{eq:unitaryinteractionU}, Eq.\ref{eq:unitaryinteractionA}, Eq.\ref{eq:unitaryinteractionVstar} and Eq.\ref{eq:unitaryinteractionUstar}, the 4-dimensional Lorentz invariant scalar, and vector operators that generate dominant contribution to baryon and lepton number violations can be obtained and are given in Table.\ref{tab:4dops}. Indeed, the $C_1^S$ and $C_2^S$ Wilson Coefficients are generated by scalar new physics, whereas, $C_1^V$ and $C_2^V$ are generated by vectors. For simplicity, we have kept only the dominant term containing zero mode of doublet Standard Model fermions among these 4-dimensional operators. The rest of the contribution can be derived similarly. The dimension-8 operators require additional quartic interactions of scalar bosons but we do not get into the details here.

\begin{table}[h!]
\centering
\begin{tabular}{|c|c|c|}
\hline 

operators & $\Delta B=1=\Delta L$ & $\Delta B=2=\Delta L$ \\
[1ex] 
\hline
&& \\
scalar & $ \frac{C^S_1}{\Lambda_4^2}(q^{(0,0)T}_{+l}\gamma^2\gamma^0\mathcal{U}_{-l}^{(1,0)})(\mathcal{E}_{-l}^{T(1,0)}\gamma^2\gamma^0 q_{+l}^{(0,0)})$ & $\frac{C^S_2}{\Lambda_4^{8}}(q^{(0,0)T}_{+l}\gamma^2\gamma^0\mathcal{U}_{-l}^{(1,0)})^2(\mathcal{E}_{-l}^{T(1,0)}\gamma^2\gamma^0 q_{+l}^{(0,0)})^2 $ \\[1ex] 
vector & $\frac{C^V_1}{\Lambda_4^2}(u^{(0,0)T}_{-r}\gamma^2\gamma^0 \gamma^\mu \mathcal{D}^{(1,0)}_{-l})(u^{(0,0)T}_{-r}\gamma^2\gamma^0 \gamma_\mu \mathcal{E}^{(1,0)}_{-l})$ & $\frac{C^V_2}{\Lambda_4^{8}}(u^{(0,0)T}_{-r}\gamma^2\gamma^0 \gamma^\mu \mathcal{D}^{(1,0)}_{-l})^2(u^{(0,0)T}_{-r}\gamma^2\gamma^0 \gamma_\mu \mathcal{E}^{(1,0)}_{-l})^2 $   \\[1ex] \hline
\end{tabular}
\caption{The list of dominant four-dimensional Lorentz invariant baryon number violating operators, that originate from the interactions in Eq.\ref{eq:scalarint}, Eq.\ref{eq:unitaryinteractionU} - Eq.\ref{eq:unitaryinteractionUstar}.}
\label{tab:4dops}
\end{table}
On the other hand, using the interactions of `spinless' adjoint partner of the hypercharge boson, given in Eq.\ref{eq:DMqint} and Eq.\ref{eq:DMeint}, the above operators generate operators given in Table.\ref{tab:4dopsVV}. 
\begin{table}[h!]
\centering
\begin{tabular}{|c|c|}
\hline 

operators & $\Delta B=1=\Delta L$  \\
[1ex] 
\hline
& \\
scalar & $ y_u y_e g_Y^2 \frac{C^S_1}{\Lambda_4^{2}}\frac{1}{M_{KK}^2}(q^{T(0,0)}_{+l} \gamma^2\gamma^0 u^{(0,0)}_{-r})(e^{T(0,0)}_{-r} \gamma^2\gamma^0 q^{(0,0)}_{+l}) V_B^{(1,0)} V_B^{(1,0)}$\\[1ex] 
vector & $y_u y_e g_Y^2 \frac{C^V_1}{\Lambda_4^2}\frac{1}{M_{KK}^2}(u^{(0,0)T}_{-r}\gamma^2\gamma^0 \gamma^\mu d^{(0,0)}_{-r})(u^{(0,0)T}_{-r}\gamma^2\gamma^0 \gamma_\mu e^{(0,0)}_{-r})V_B^{(1,0)} V_B^{(1,0)}$  \\[1ex] \hline
operators & $\Delta B=2=\Delta L$  \\
[1ex] 
\hline
& \\
scalar & $ y_u^2 y_e^2 g_Y^4 \frac{C^S_2}{\Lambda_4^{8}}\frac{1}{M_{KK}^4}(q^{(0,0)T}_{+l}\gamma^2\gamma^0 u_{-r}^{(0,0)})^2(e_{-r}^{T(0,0)}\gamma^2\gamma^0 q_{+l}^{(0,0)})^2 V_B^{(1,0)} V_B^{(1,0)}V_B^{(1,0)} V_B^{(1,0)}$\\[1ex] 
vector & $y_u^2 y_e^2 g_Y^4 \frac{C^V_2}{\Lambda_4^8}\frac{1}{M_{KK}^4}(u^{(0,0)T}_{-r}\gamma^2\gamma^0 \gamma^\mu d^{(0,0)}_{-r})^2(u^{(0,0)T}_{-r}\gamma^2\gamma^0 \gamma_\mu e^{(0,0)}_{-r})^2V_B^{(1,0)} V_B^{(1,0)}V_B^{(1,0)} V_B^{(1,0)}$  \\[1ex] \hline
\end{tabular}
\caption{The baryon number violating operators with Standard Model zero mode fermions in the external legs, generated after including the interaction of `spinless' adjoint scalar field, given in Eq.\ref{eq:DMqint} and Eq.\ref{eq:DMeint}.}
\label{tab:4dopsVV}
\end{table}
We will analyse the phenomenology of these operators in the coming sections.

\section{(Assisted) Proton decay}
\label{sec:Aprotondecay}

After orbifolding and integrating out the extra-dimensions, the operators that contribute to the baryon number violation by one unit, in Table.\ref{tab:4dops}, are
\begin{eqnarray}
  \mathcal{O}_{1}&=& \dis \frac{C^S_1}{\Lambda_4^2}(q^{(0,0)T}_{+l}\gamma^2\gamma^0\mathcal{U}_{-l}^{(1,0)})(\mathcal{E}_{-l}^{T(1,0)}\gamma^2\gamma^0 q_{+l}^{(0,0)}) +  \frac{C^V_1}{\Lambda_4^2}(u^{(0,0)T}_{-r}\gamma^2\gamma^0 \gamma^\mu \mathcal{D}^{(1,0)}_{-l})(u^{(0,0)T}_{-r}\gamma^2\gamma^0 \gamma_\mu \mathcal{E}^{(1,0)}_{-l})  \ . \nonumber \\
\label{eq:peo4d}
\end{eqnarray}
The scalar operator generates exotic baryon number violating current, where KK-1 partner of the right-handed up quark decays to two SM quarks and a lepton KK-1 mode. Such decays are allowed due to the $\sim 20 $ GeV mass split~\cite{Cheng:2002iz, Ponton:2005kx, Dobrescu:2007ec} between the KK-1 modes of up quark and the lepton~\cite{Cheng:2002iz}. 

\begin{figure}
\begin{subfigure}{0.4\textwidth}
    \includegraphics[width=\textwidth]{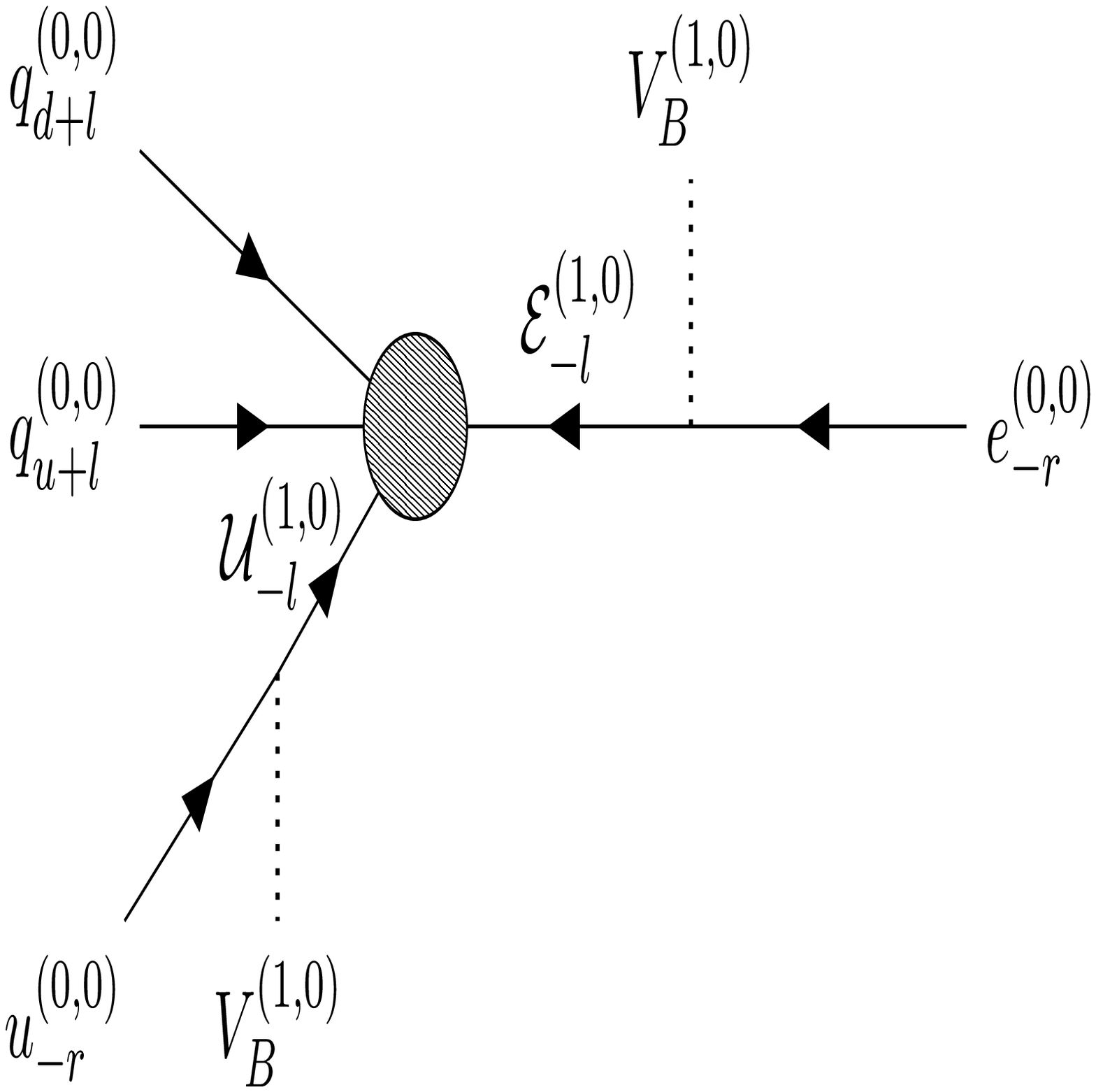}
\vspace {-3cm}
\caption{}
\label{fig:peVV}
\end{subfigure}
\begin{subfigure}{0.4\textwidth}
    \includegraphics[width=\textwidth]{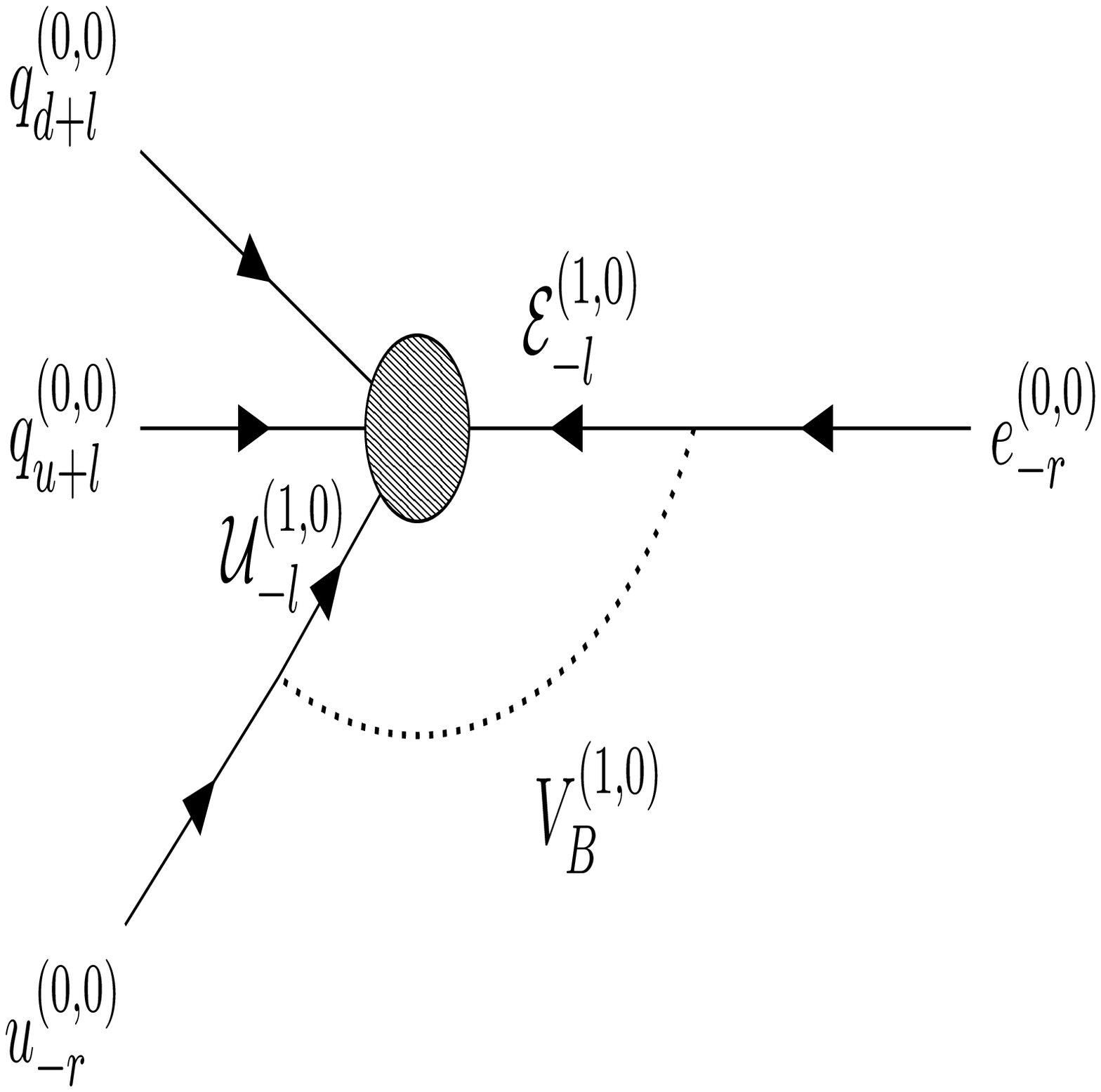}
\vspace {-3cm}
\caption{}
\label{fig:peloop}
\end{subfigure}
\vspace {0cm}
\caption{Figure (a) shows the process $DM + p \rightarrow DM +  e^+ + \pi^0$ generated through the scalar operator in Eq.\ref{eq:AND}. The stable spinless adjoint scalar field, $V_B^{(1,0)}$, is the Dark Matter candidate. Figure (b) shows the process $p \rightarrow  e^+ + \pi^0$ generated at 1-loop.}
\label{fig:AND}
\end{figure}

Along with the adjoint scalar interactions, the first term in Eq.\ref{eq:peo4d} becomes,
\begin{eqnarray}
    C_{AND} \mathcal{O}_{AND} &=& \dis  y_u y_{e} g_Y^2 \frac{C^S_1}{\Lambda_4^{2}}\frac{1}{M_{KK}^2}(q^{T(0,0)}_{+l} \gamma^2\gamma^0 u^{(0,0)}_{-r})(e^{T(0,0)}_{-r} \gamma^2\gamma^0 q^{(0,0)}_{+l}) V_B^{(1,0)} V_B^{(1,0)} \ ,
\label{eq:AND}
\end{eqnarray}
where $g_Y$ is the coupling for the KK-1 hypercharge spinless adjoint scalar field and $y_{u}=4/3$, $y_{e}=-2$. The processes generated by this operator are shown in Fig.\ref{fig:AND}. With SM like interactions, the hypercharge coupling is given by $g_Y^2 = \frac{4 M_z^2 Sin\theta_W^2}{ v^2} \simeq 0.14 $. This operator leads to assisted proton decay process in the early epochs of the universe, also provides new annihilation channel for the Dark Matter. 

Note that at 1-loop, as shown in Fig.\ref{fig:peloop}, this operator contribute to the direct proton decay. The effective operator in Eq.\ref{eq:AND} for the process, after integrating out the loop, becomes,
\begin{equation}
   C_{p\to e} \mathcal{O}_d^{(2)} =  y_{u} y_{e} g_Y^2\frac{C^S_1}{16 \pi^2 \Lambda_4^2} \Big(\frac{M_s}{M_{KK}}\Big)^4 (u^{T(0,0)}_{+l} \gamma^2\gamma^0 d^{(0,0)}_{+l})(e^{T(0,0)}_{-r}\gamma^2\gamma^0 u^{(0,0)}_{-r}) \ ,
\label{eq:protondecayVVloop}
\end{equation}
where $M_s$ is the loop momentum and the Wilson Coefficient for the decay can be read off as,
\begin{equation}
    C_{p \rightarrow e} = y_{u} y_{e} g_Y^2\frac{C_1^S}{16 \pi^2\Lambda_4^2} \Big(\frac{M_s}{M_{KK}}\Big)^4 \ .
\end{equation}

After matching the quark level operator in Eq.\ref{eq:protondecayVVloop} with the nucleon decay matrix element using $\chi PT$~\cite{JLQCD:1999dld}, the hadronic operator generates the decay width,
\begin{equation}
    \Gamma_{p\to e} = \frac{1}{2 \times 10^{34}} \Big|\frac{C_{p \to e}}{(3 \times 10^{15} \ {\rm GeV})^{-2}}\Big|^2 \ .
    \label{eq:ptoe}
\end{equation}

Then, from the above relation, assuming $C^S_1 =0.01$, $M_{KK}=10 $ TeV and $M_s = m_p$, the New Physics that contributes to the proton decay can be constrained to be $\gtrsim 140 $ TeV. This scale can be relaxed further to $\sim 40 $ TeV, which is within the reach of possible future 100 TeV hadron collider~\cite{Mangano:2017tke} for Wilson Coefficient $C_1^S \sim \mathcal{O}(10^{-3})$. 

Getting back to the assisted nucleon decay (AND), in the scenario in which the `spinless' adjoint scalar becomes DM, the effective rate can be written as,
\begin{equation}
\Gamma_{AND} = n_{DM} (\sigma v)_{AND} \ ,
\end{equation}
where $n_{DM}$ is the Dark Matter density and $(\sigma v)_{AND}$ is the cross section for the assisted proton decay process. Using Eq.\ref{eq:AND}, the assisted proton decay cross section can be computed as,
\begin{equation}
(\sigma v)_{AND} \sim \frac{1}{16 \pi} \Big| \frac{ y_{u} y_e g_Y^2 C_1^S}{\Lambda_4^2 M_{KK}^2}  \Big|^2 m_p^6 \ .
\end{equation}
The lifetime of this process is then given by,
\begin{equation}
\tau_{AND} = \frac{1}{\Gamma_{AND}} = \frac{M_{KK}}{\rho_{DM} (\sigma v)_{AND}} \ ,
\label{eq:ANDlifetime}
\end{equation}
where, the number density of DM has been replaced with the mass density $\rho_{DM} = M_{KK} n_{DM} = 0.3 GeV/cm^3$. Using the scale of the operator previously computed, $\Lambda_4 \sim 140 $ TeV, the time period for the assisted proton decay, given in Eq.\ref{eq:ANDlifetime} and assuming $C_1^S = 0.01$, becomes $\tau_{AND} \gg 1.4 \times 10^{34} $ years, satisfying the constraint from Super-Kamiokande~\cite{Super-Kamiokande:2018apg}. Thus, the model suggests that the rareness of Dark Matter density on Earth results in the assisted nucleon decay time period much beyond the observational sensitivity of terrestrial experiments. Nevertheless, note that, this process produces a striking signature with highly collimated pion and positron Cherenkov rings. Moreover, the assisted proton decay can be much more enhanced near large Dark Matter densities like center of the galaxy. And since the process preserves the Dark Matter number density, this will play a major role in the baryon number violation near very heavy astrophysical objects.

\section{(Assisted) $\Delta B=2$, $\Delta L=2$ process}
\label{sec:B2L2}

\begin{figure}
\centering
\begin{subfigure}{0.45\textwidth}
    \includegraphics[width=\textwidth]{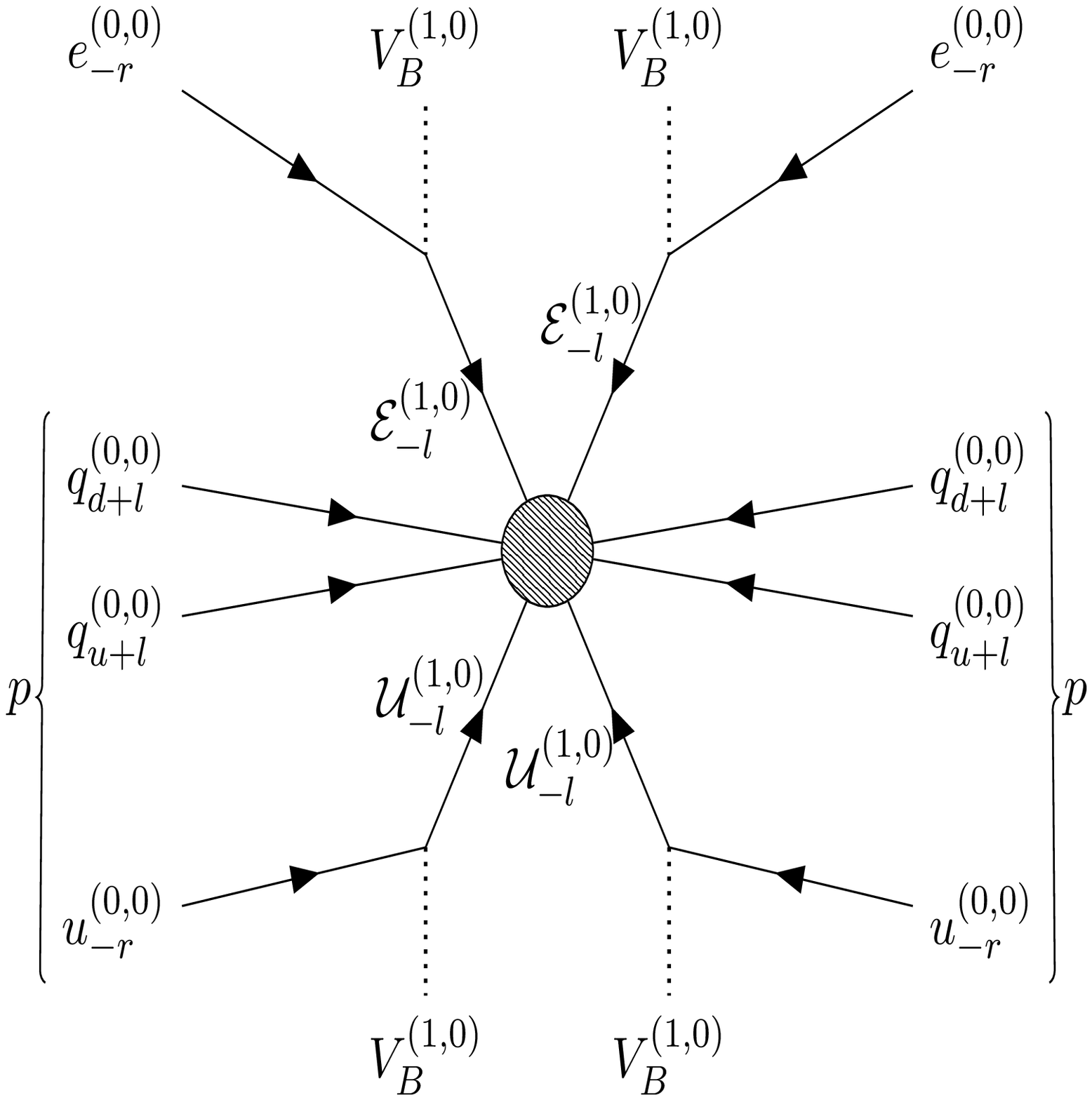}
\vspace {-3cm}
\caption{}
\label{fig:peVV}
\end{subfigure}
\begin{subfigure}{0.45\textwidth}
    \includegraphics[width=\textwidth]{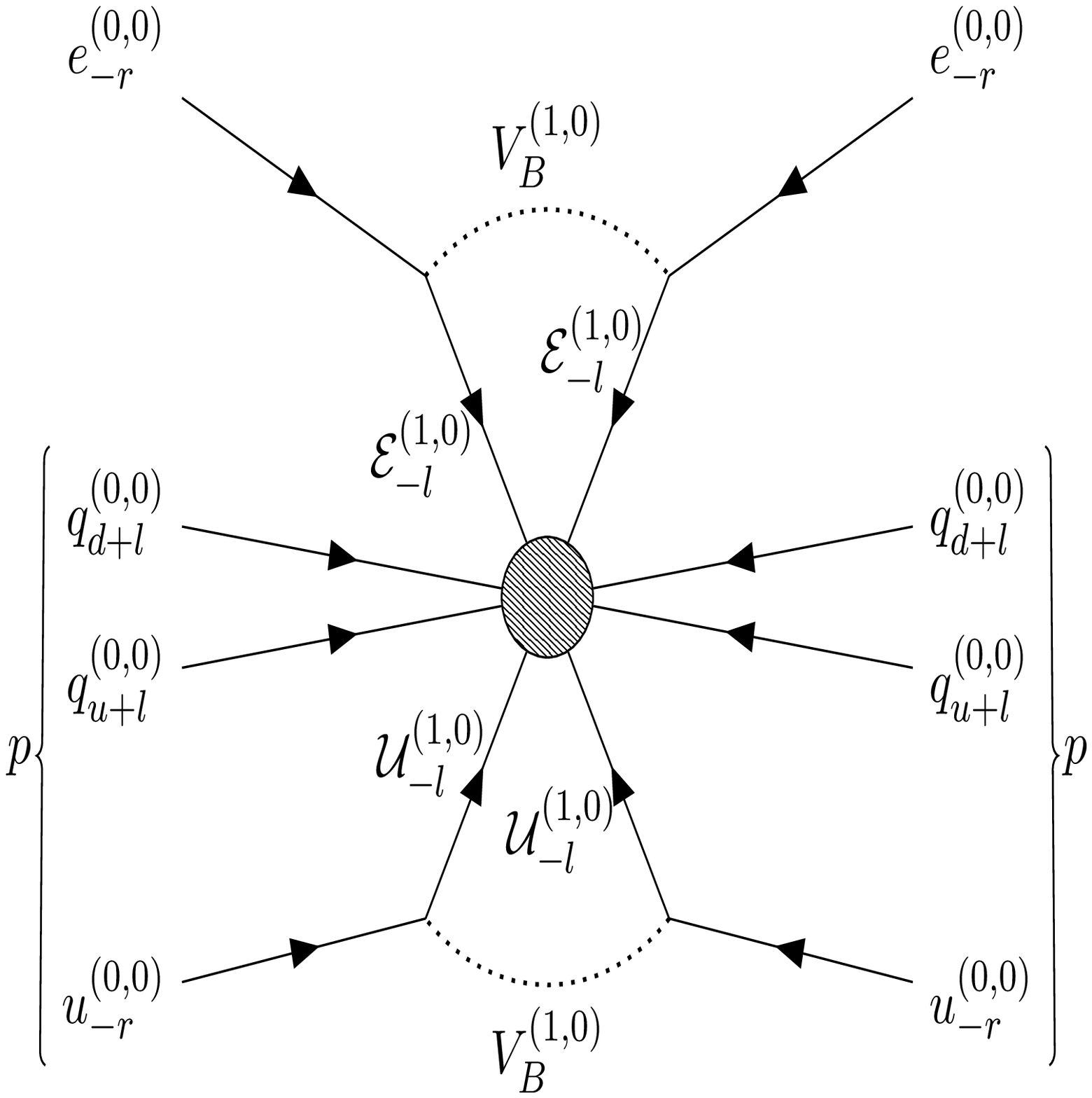}
\vspace {-2.8cm}
\caption{}
\label{fig:peloop}
\end{subfigure}
\vspace {0cm}
\caption{Figure (a) shows the processes $\Delta B=2$, $\Delta L=2$ given in Eq.\ref{eq:ANNA}. The stable adjoint scalar field, $V_B^{(1,0)}$, is the Dark Matter candidate. Figure (b) shows the process generated at 2-loop.}
\label{fig:ANNA}
\end{figure}

After orbifolding and integrating out the extra-dimensions, the operators that contribute to violation of baryon number and lepton number by two units, given in Table.\ref{tab:4dops}, are,
\begin{eqnarray}
    \mathcal{O}_{2} &=& \dis  \frac{C^S_2}{\Lambda_4^{8}}(q^{(0,0)T}_{+l}\gamma^2\gamma^0\mathcal{U}_{-l}^{(1,0)})^2(\mathcal{E}_{-l}^{T(1,0)}\gamma^2\gamma^0 q_{+l}^{(0,0)})^2 \nonumber \\
    && \dis+ \frac{C^V_2}{\Lambda_4^{8}}(u^{(0,0)T}_{-r}\gamma^2\gamma^0 \gamma^\mu \mathcal{D}^{(1,0)}_{-l})^2(u^{(0,0)T}_{-r}\gamma^2\gamma^0 \gamma_\mu \mathcal{E}^{(1,0)}_{-l})^2  \ . 
\label{eq:4dHHops}
\end{eqnarray} 
These operators, along with the `spinless' adjoint scalar interaction term given in Eq.\ref{eq:DMqint}, generates the assisted nucleon nucleon annihilation (ANNA). An example of this process, generated from the scalar operator, is given by the effective Lagrangian term,
\begin{eqnarray}
    C_{ANNA}\mathcal{O}_{ANNA}
    &=& \dis y_{u}^2y_{e}^2 g_Y^4 \frac{C^S_{2}}{\Lambda_4^{8} M_{KK}^4}(q^{(0,0)T}_{+l}\gamma^2\gamma^0 u_{-r}^{(0,0)})^2(e_{-r}^{T(0,0)}\gamma^2\gamma^0 q_{+l}^{(0,0)})^2 V_{B}^{(1,0)}V_{B}^{(1,0)}V_{B}^{(1,0)}V_{B}^{(1,0)} \ . \nonumber \\
\label{eq:ANNA}
\end{eqnarray} 
If the adjoint scalar becomes the Dark Matter, this interaction generates $DM+p + DM + p \to DM + e^+ + DM + e^+$, and assisted Hydrogen-antiHydrogen oscillation as shown in Fig.\ref{fig:ANNA}(a). Though the probability of this process is very small on Earth, they can be substantial near Dark Matter clumps. These processes can be interesting to study in the context of the observed positron excess in cosmic rays~\cite{PAMELA:2008gwm, Delahaye:2008ua, Fermi-LAT:2011baq, PAMELA:2013jtv, PAMELA:2013vxg, DeSarkar:2019tjy}.

At loop level there exists processes like $p+p \to e^+ + e^+$, $DM+ p \to DM+ \bar{p}+ e^+ + e^+$ and $(DM+ e)+ p \to (DM+ e^+)+ \bar{p}$, but except for the first, current experiments do not constraint the rest directly. Nevertheless, we re-derive the bounds on these processes using the results from $pp\to e^+e^+$ below.

\subsection{$p+p \to e^+ + e^+$}
At 2-loop, the diagram given in Fig.\ref{fig:peloop} generates processes like Hydrogen-antiHydrogen oscillation and $p + p \to \ell^+ + \ell^+$, where $\ell^+ = (e^+,\mu^+)$. For the proton proton annihilation process, the hadron-level effective operator becomes,
\begin{equation}
\frac{1}{\Lambda_{ppee}^2}(\bar{p}^c \gamma^5p)(\bar{e}^c \gamma^5 e) \ .
\label{eq:ppee}
\end{equation}
At Super-Kamiokande, this term is constrained by studying the process ${}^{16} O \rightarrow {}^{14}C + \ell^+ + \ell^+$, with same-sign dilepton back-to-back Cherenkov rings with no hadrons. With a fiducial mass of 22.5 kilotons containing $\sim 10^{34}$ nucleons, the time period of the decay per oxygen nucleus is constrained to $T_{pp\to e^+ e^+} = 4.2 \times 10^{33} $ years~\cite{Super-Kamiokande:2018apg}.
Comparing Eq.\ref{eq:ppee} with the first term in Eq.\ref{eq:4dHHops}, and using the well known nuclear matrix element, we get,
\begin{equation}
    \frac{1}{\Lambda_{ppee}^2} = \frac{C_2^S}{\Lambda_4^{8}} *(0.22 m_p)^6 \Big(\frac{1}{16 \pi^2}\Big)^2 \Big(\frac{Ms}{M_{KK}}\Big)^4 \ .
\end{equation}
Assuming a New Physics scale $\Lambda_4 \sim 2 $ TeV, with $C_2^S \sim \mathcal{O}(1)$ and $M_{KK} = 10 $ TeV, the  hadronic effective operator gets highly suppressed beyond the current sensitivity of experiments.

Whereas, the process $p e \rightarrow \bar{p} e^+$, obtained by Fierz transforming the relation in Eq.\ref{eq:ppee} can lead to Hydrogen-antiHydrogen oscillation, studied by measuring the $\gamma$ rays from the annihilation of antiHydrogen at the interstellar medium~\cite{Grossman:2018rdg}, and annihilation of proton in a nucleus with an electron in the inner shell of oxygen at Super-Kamiokande~\cite{Super-Kamiokande:2018apg}.

\subsection{$DM+ p \to DM+ \bar{p}+ e^+ + e^+$}
Interestingly, the term in Eq.\ref{eq:4dHHops} generate baryon number violating processes in which Dark Matter interacts with the proton legs. The most constraining among these processes is the one in which a Dark Matter scatters with one proton producing same sign dileptons, an anti-proton ($\bar{p}$) and Dark Matter at rest ($ DM+ p \rightarrow DM+ \bar{p} + e^+ + e^+$). With the $\bar{p}$ annihilating with a another proton, the results of $p+p\to e^+ + e^+$ can be recasted here. Since this is initiated by a heavy Dark Matter, the dileptons produced will be highly energetic. They can be searched for at Super-Kamiokande through two collinear Cherenkov rings. To analyse this, lets write the hadron-level operator,
\begin{eqnarray}
\mathcal{O}_{VVppee} &=& \dis \frac{1}{(\Lambda_{VVppee})^4}(\bar{p}^c p)(\bar{e}^c e) V_B^{(1,0)} V_B^{(1,0)} \ ,
\label{eq:VpVpee}
\end{eqnarray}
where, comparing with Eq.\ref{eq:ANNA}, we get,  
\begin{eqnarray}
\frac{1}{(\Lambda_{VVppee})^4} &=& \dis y_u y_eg_Y^2 \frac{C^S_{2}}{\Lambda_4^{8}} \frac{1}{M_{KK}^2}(0.22 m_p)^6 \frac{1}{16\pi^2} \Big(\frac{M_s}{M_{KK}}\Big)^2 \ .
\end{eqnarray}
The effective decay width for the nucleon, then, becomes,
\begin{equation}
\Gamma_{ANNA} = n_{DM} (\sigma v)_{ANNA} \ ,
\end{equation}
where $n_{DM}$ is the Dark Matter density and $(\sigma v)_{ANNA} \sim \frac{1}{16 \pi} \Big| \frac{1}{(\Lambda_{VVppee})^4} \Big|^2 m_p^6 $ is the cross section for the assisted double nucleon decay process given by the operator in Eq.\ref{eq:VpVpee}. The lifetime of this process is then given by,
\begin{equation}
\tau_{ANNA} = \frac{1}{\Gamma_{ANNA}} = \frac{M_{KK}}{\rho_{DM} (\sigma v)_{ANNA}} \ .
\end{equation}
In the above equation, we have replaced the number density of DM with the mass density $\rho_{DM} = M_{KK} n_{DM} = 0.3 GeV cm^{-3}$. For Dark Matter of mass $M_{KK}=2 $ TeV, from galactic center, with speed $\sim 100 km/s$, colliding with an ${}^{16}O$ atom in the experiment, the average transfer momentum can be computed to be $P_t = m_{V} v \sim 600 $ MeV. In this process, $DM + p \rightarrow DM + \bar{p} + e^+ + e^+ $, since $m_e \ll P_t < m_p \ll m_V$, we can safely assume that the DM and anti-proton are produced at rest. Thus, for all practical considerations, this is a $2 \rightarrow 2$ process with proton in the incoming leg at rest in the Lab frame. 

Using the limit on double proton decay time period $\tau_{pp\to e^+e^+} \gtrsim 4.2 \times 10^{33} years $ ~\cite{Super-Kamiokande:2018apg}, we get, 
\begin{equation}
\tau_{ANNA} = 5 \times 10^{33} years \Big( \frac{\Lambda_{VVppee}}{ 300 \ {\rm GeV}} \Big)^8 \ .
\end{equation}
This is a very weak limit for the New Physics model, thus terrestrial experiments are not very sensitive to the the assisted nucleon nucleon decay yet. On the other hand, the clean and unique signal for this event is very interesting, in case a Dark Matter interacts in the upcoming Hyper-Kamiokande~\cite{Hyper-Kamiokande:2018ofw} experiment. The constraint on this operator is weak due to the rarity of Dark Matter density on Earth, whereas, in primordial Superdense cosmological dark matter clumps~\cite{PhysRevD.81.103529,PhysRevD.81.103530} with large gravitating mass, this may not be the case. Such assisted double nucleon decays can be a very large source of baryon number violation in cosmology.
The processes such as assisted Hydrogen oscillation ($ DM (p e) \rightarrow DM (\bar{p} e^+)$), can also be searched for at the interstellar medium by Fermi-Lat and can give complementary measurements for the operator.

This process also generates baryon number violation in which the Dark Matter interacts with the lepton legs. Though, this process is kinematically prohibited at low energies, they can be probed in high energy collider experiments. Such processes can be constrained by the CMS~\cite{CMS:2017tec} study where they consider finals states with two same sign leptons, two or more hadronic jets and missing energy. Unfortunaltey, since the operator is at mass dimension-10, it is highly suppressed and moreover, at high energy collider, the patrons will probe the insides of the effective operator. Though the best constraint on this operator might arise from the collider experiments, this will not be a model independent result.

\subsection{$(DM+e) + p \to (DM+ e^+) + \bar{p} $}

On the other hand, $(DM+e) + p \to (DM+ e^+) + \bar{p} $, or DM assisted Hydrogen-antiHydrogen oscillation can be a better probe to study this operator. The relevant constraint emanates from the non-obeservation of Hydrogen-antiHydrogen annihilation $\gamma-$rays from Inter Stellar Medium (ISM) surveyed by Fermi LAT. 

We consider the scenario where the Hydrogen atoms, in its ground state, is influenced by Dark Matter.  Then the oscillation Hamiltonian,
\begin{equation}
H_{osc} = \frac{1}{\Lambda_{AH\bar{H}}^4} (\bar{p}^c e )(\bar{p}^c  e)V_{B}^{(1,0)}V_{B}^{(1,0)} \ ,
\end{equation}
generates small amounts of antiHydrogen. Since the conversion is enabled by Dark Matter number density ($n_{DM}$) in the ISM, the rate of antiHydrogen production is given by,
\begin{eqnarray}
\tau_{osc} = \frac{1}{\Gamma_{osc}}= \frac{1}{n_{DM} (\sigma v)_{osc}} \ ,
\end{eqnarray}
where, $(\sigma v)_{osc} = \frac{1}{16 \pi} \Big(\frac{m_p^6}{\Lambda_{AH\bar{H}}^8}\Big) \ ,$
is the cross section for the process. 

Then, the rate of assisted Hydrogen oscillation can be computed by studying the production rate of $\gamma$ rays due to the annihilation of the antiHydrogen with Hydrogen~\cite{PhysRevD.18.1602}. For a New Physics, presumably at $\Lambda_{AH\bar{H}}$ of $\mathcal{O}(1\rm{ TeV } )$, the width of the process can be computed to be $\Gamma_{osc} \sim 7 \times 10^{-44} s^{-1}$. Thus, the constrain placed by the analysis of $\gamma$ ray data from Fermi LAT in the range $100 MeV - 9.05 $ GeV~\cite{Grossman:2018rdg} does not constraint this operator. More reliable bound can be obtained from 14 TeV or 100 TeV LHC, but that requires a UV complete model. Nevertheless, this process also depends on the Dark Matter density. Thus, galaxy cluster centers can be a good source of assisted Hydrogen-antiHydrogen oscillation.  

\section{Conclusion}
\label{sec:conclusion}

The phenomenology in $4k+2-$dimensions, in particular six-dimensions is very interesting. Vanishing of Witten anomaly in six-dimensions, arising from the non-trivial winding of the spacetime on $SU(2)_W$ given by $\pi_6(SU(2)_W) = Z_{12}$, correctly predicts the  fermion generations charged under $SU(2)_w$ gauge group to be multiples of 3. The gauge bosons in dimensions $d \geq 6$ also exhibit interesting properties. In an uncompactified geometry, these gauge bosons have $d-2$ polarization vectors. Upon compactification, $d-4$ of them break and one combination among them is ``eaten'' by the KK modes making them heavy. The rest of the broken polarizations become the `spinless' adjoint scalar fields and remain in KK spectrum with mass $\sim 1/R $, where $R$ is the compactification radius. In general, due to mass corrections, at 1-loop, the degeneracy among all the lightest KK masses are lifted leading to a mass hierarchy, and the `spinless' adjoint scalar of the hypercharge gauge boson becomes the lightest. Including the KK-partiy conservation, this KK-1 scalar is stable and hence is the Dark Matter candidate in the model. The relic density constraint places a bound of $\sim 2 $ TeV on the compactificaitons scale, assuming that the adjoint scalar makes up the entire Dark Matter density. This limit can be relaxed by introducing additional resonant annihilation and co-annihilation channels, either by including additional fields or embedding the six-dimensions in a seven-dimensional space-time with only gravity allowed to propagate in the bulk~\cite{Arun:2017zap}. 

The spinor properties in $4k+2-$dimension are different from $4k-$dimension. In $4k+2$ dimensions, the charge conjugation operator commutes with the chiral projection operator, thus keeping the chirality unchanged under charge conjugation of the fermion. And its consequences are interesting in the context of baryon number and lepton number violating currents. In this article, we analyse such effective operators in six-dimensions, as a minimal extension. For simplicity and clarity, we explicitly work with Standard Model and New Physics (scalar and vector bosons) in the bulk. Though these New Physics fields, in four-dimensions, generate the usual baryon number violation, in orbifolded six-dimensions they generate novel operators given in Table.\ref{tab:4dops}. In six-dimensions, the tree-level proton decay operator with only KK zero modes in their external legs have been shown to be highly suppressed~\cite{Appelquist:2001mj} due to the emergent selection criteria $3/2\Delta B \pm 1/2 \Delta L = 0 \ mod \ 4$. Here, we show that this selection criteria can be circumvented if we allow higher Kaluza Klein modes in the operators. Moreover, these novel operators generate Dark matter assisted baryon number violating processes at mass dimension-8 and higher upon including the interaction of the `spinless' adjoint scalar field. These processes, among others, contain assisted proton decay $V_B^{(1,0)} + p \to V_B^{(1,0)} + e^+ + \pi^0$ and assisted Hydrogen-antiHydrogen oscillation $V_B^{(1,0)} + p + e \to V_B^{(1,0)} + \bar{p} + e^+ $. We show that the proton decay data from SuperKamiokande constraints the assisted proton decay operator to $\gtrsim 1400 $ TeV, for $1/R = 10 $ TeV and $C_1^S = \mathcal{O}(1)$. Though the kinematics of this $2 \to 3$ process is different, it can be identified in the water Cherenkov detector with rings corresponding to a positron and a pion~\cite{Huang:2013xfa}. The lower bound on the New Physics can indeed be brought within the reach of 100TeV collider in the weak coupling limit, $C^S_1 \lesssim 10^{-3}$. But in the upcoming HyperKamiokande experiment, with its better sensitivity to the proton decay process, the scale of New Physics could be constrained by a factor $\mathcal{O}(10)$. 

Unlike the models with baryon number violating scalar and vector bosons in four-dimensions, here we show that the rarity of the processes at terrestrial experiments could be explained by the lack of enough Dark Matter density on Earth. On the other hand, they have larger probability to occur near Dark Matter clusters and can be uniquely identified by studying the positron fluxes emerging from such clusters. Such a scenario could also arise from Dark Matter accumulation in Sun.

\acknowledgments
M.T.A. acknowledges the financial support of DST through INSPIRE Faculty grant 
 DST/INSPIRE/04/2019/002507. 

\section{Appendix}
\label{appendix:a}

\subsection{Standard Model hypercharge gauge field and its KK modes}
\label{section:A3}
The Lagrangian density of an Abelian gauge field in six dimensions is given by,
\begin{equation}
    \mathcal{L}=\frac{1}{4} F^{MN}F_{MN}+ \mathcal{L}_{GF} \ ,
\end{equation}
where $M=0,1,2,3,4,5$ and $\mathcal{L}_{GF}$ is the appropriate gauge fixing term. 
Since the six-dimensional geometry is orbifolded on $T^2/Z_2$, we use the generalised $R_{\xi}$ gauge,
\begin{equation}
\mathcal{L}_{GF}=-\frac{1}{2\xi} (\partial_\mu \mathcal{A}_\mu - \xi (\partial_4 \mathcal{A}_4+\partial_5 \mathcal{A}_5))^2 \ .
\end{equation}

Expanding the terms we get,
\begin{eqnarray}
  \mathcal{L}&=&\frac{1}{4} F^{\mu \nu}F_{\mu \nu}+\frac{1}{2}((\partial_\mu \mathcal{A}_4- \partial_4 \mathcal{A}_\mu)^2+ (\partial_\mu \mathcal{A}_5- \partial_5 \mathcal{A}_\mu)^2+ (\partial_5 \mathcal{A}_4- \partial_4 \mathcal{A}_5)^2)
  -\frac{1}{2\xi}(\partial_\mu \mathcal{A}_\mu)^2 \nonumber \\ &-&\frac{\xi}{2}( (\partial_4 \mathcal{A}_4)^2+(\partial_5 \mathcal{A}_5)^2 + 2(\partial_4 \mathcal{A}_4)(\partial_5 \mathcal{A}_5))-(\partial_\mu \mathcal{A}_\mu) (\partial_4 \mathcal{A}_4) -(\partial_\mu \mathcal{A}_\mu)(\partial_5 \mathcal{A}_5) \ .
  \label{4dleg}
\end{eqnarray}

Upon orbifolding, the $\mathcal{A}_{\mu}(x^{\mu},x^4,x^5)$ is set to satisfy Neumann boundary condition and $\mathcal{A}_{4}(x^{\mu},x^4,x^5), \ \mathcal{A}_{5}(x^{\mu},x^4,x^5)$ are set to satisfy Dirichlet boundary condition at both the brane positions. Hence, they get decomposed in terms of their KK modes as,
\begin{eqnarray}
\mathcal{A}_{\mu}(x^{\mu},x^4,x^5)&=& \mathcal{A_{\mu}}^{(0,0)}(x^{\mu})+\sqrt{2} \sum_{m,n}
\mathcal{A}^{(m,n)}_{\mu}(x^{\mu})cos\left[\frac{1}{R}(mx_4+n x_5) \right]\nonumber, \\
\mathcal{A}_4(x^{\mu},x^4,x^5)&=& \sum_{m,n}
\mathcal{A}^{(m,n)}_4(x^{\mu})sin\left[\frac{1}{R}(mx_4+n x_5) \right] \nonumber \\
\mathcal{A}_5(x^{\mu},x^4,x^5)&=& \sum_{m,n}
\mathcal{A}^{(m,n)}_5(x^{\mu})sin\left[\frac{1}{R}(mx_4+n x_5) \right]
\label{eq:SMgaugeKK}
\end{eqnarray}
The zero mode of $\mathcal{A}_\mu(x^\mu,x^4,x^5)$ is identified with the 4-dimensional Standard Model gauge field, while $\mathcal{A}_4$ and $\mathcal{A}_5$ are adjoint scalar fields that arise from the two broken polarisations of the six-dimensional gauge field. Note that these adjoint scalar fields do not posses zero modes. 

After integrating out the $x_4, \ x_5$ coordinates, the partial derivatives $\partial_4, \partial_5$
is replaced by $\frac{m}{R}$ and $\frac{n}{R}$ respectively. The simplified 4-dimensional Lagrangian density is given by,
\begin{eqnarray}
  \mathcal{L}&=& \dis \frac{1}{4} F^{(m,n)\mu \nu}F^{(m,n)}_{\mu \nu}-\frac{1}{2\xi}(\partial_\mu \mathcal{A}^{(m,n)}_\mu)^2 + (\partial_\mu \mathcal{A}^{(m,n)}_4)^2+(\partial_\mu \mathcal{A}^{(m,n)}_5)^2 \nonumber\\
&+& \dis \frac{1}{2}(\frac{n}{R} \mathcal{A}^{(m,n)}_4- \frac{m}{R} \mathcal{A}^{(m,n)}_5)^2  \nonumber\\
&-& \dis \frac{\xi}{2 R^2}(m^2 \mathcal{A}^{(m,n)2}_4+n^2 \mathcal{A}^{(m,n)2}_5+ 2 mn \mathcal{A}^{(m,n)}_4 \mathcal{A}^{(m,n)}_5)
\end{eqnarray}
In the above equation, the Lagrangian term with $\frac{\xi}{2}$ can be written in matrix form,
\begin{equation}
 \mathcal{L_\xi}=\begin{bmatrix}\mathcal{A}^{(m,n)}_{4}&\mathcal{A}^{(m,n)}_{5}\end{bmatrix}\frac{1}{R}\begin{bmatrix}
m^2&mn\\mn&n^2\end{bmatrix}\frac{1}{R}\begin{bmatrix}\mathcal{A}^{(m,n)}_{4}\\\mathcal{A}^{(m,n)}_{5}
\end{bmatrix}
\end{equation}
After diagonalising this matrix we get,
\begin{equation}
  \mathcal{L_\xi}=  \begin{bmatrix}V_{1}^{(m,n)}&V_{2}^{(m,n})\end{bmatrix}\frac{1}{R}\begin{bmatrix}
{m^2+n^2}&0\\0&0\end{bmatrix}\frac{1}{R}\begin{bmatrix}V_{1}^{(m,n)}\\V_{2}^{(m,n)}\end{bmatrix} \ ,
\end{equation}
where $V_1^{(m,n)}$ and $V_2^{(m,n)}$ are given by,
\begin{eqnarray}
  V_1^{(m,n)}= \frac{m}{\sqrt{m^2+n^2}}\mathcal{A}^{(m,n)}_4+\frac{n}{\sqrt{m^2+n^2}}\mathcal{A}^{(m,n)}_5\nonumber \\
  V_2^{(m,n)}=\frac{-n}{\sqrt{m^2+n^2}}\mathcal{A}^{(m,n)}_4+\frac{m}{\sqrt{m^2+n^2}}\mathcal{A}^{(m,n)}_5
  \label{adj}
\end{eqnarray}
The fields $\mathcal{A}^{(m,n)}_4$ and $\mathcal{A}^{(m,n)}_5$ are replaced by its scalar adjoint $V_1^{(m,n)}$ and $V_2^{(m,n)}$ in Eq.(\ref{4dleg}),
\begin{eqnarray}
    \mathcal{L}&=& \dis \frac{1}{4} F^{(m,n)\mu \nu}F^{(m,n)}_{\mu \nu}-\frac{1}{2\xi}(\partial_\mu \mathcal{A}^{(m,n)}_\mu)^2 \nonumber \\
&+& \dis (\partial_\mu V^{(m,n)}_1)^2+\frac{\xi}{2 R^2}(m^2+n^2) V_1^{(m,n)2} \nonumber \\
&+& \dis (\partial_\mu V^{(m,n)}_2)^2 + (m^2+n^2)V_2^{(m,n)2}
    \label{vleg}
\end{eqnarray}
Gauge invariance implies that $\xi$ must drop out from any calculation of physical observables. The limit $\xi\to \infty$, Unitary gauge, $V_1$ becomes non-dynamical.
Hence, only the adjoint $V_2^{(m,n)}$ remain as a physical spin-$0$ particle. The lightest stable `spinless' adjoint partner of hypercharge gauge boson, $V_2^{(1,0)}$, becomes the Dark Matter candidate.  We refer to it as $V_B^{(1,0)}$  through out this article.

\bibliographystyle{unsrt}

\bibliography{bib}
\end{document}